\documentclass{article}

\usepackage[pdftex]{graphicx, color}
\usepackage{geometry}
\usepackage{bm}
\usepackage{amsmath,amsthm,amssymb,amsfonts}
\usepackage{algorithm,algorithmic}
\usepackage{authblk}
\usepackage{indentfirst}
\usepackage{url}
\usepackage[title]{appendix}
\usepackage[colorlinks=true, linkcolor=blue, filecolor=blue, urlcolor=blue, citecolor=red]{hyperref}
\usepackage{multirow}
\usepackage{colortbl,array,xcolor}
\allowdisplaybreaks[2]

\makeatletter
\long\def\@makecaption#1#2{%
  \normalsize%% add this line
  \vskip\abovecaptionskip
  \sbox\@tempboxa{#1: #2}%
  \ifdim \wd\@tempboxa >\hsize
    #1: #2\par
  \else
    \global \@minipagefalse
    \hb@xt@\hsize{\hfil\box\@tempboxa\hfil}%
  \fi
  \vskip\belowcaptionskip}
\makeatother

\title{X-DC: Explainable Deep Clustering based on Learnable Spectrogram Templates}
\author[1]{Chihiro Watanabe\thanks{chihiro.watanabe.xz@hco.ntt.co.jp}} %%%
\author[1]{Hirokazu Kameoka\thanks{hirokazu.kameoka.uh@hco.ntt.co.jp}}
\affil[1]{{\normalsize NTT Communication Science Laboratories, Kanagawa Pref. Japan}}
\date{}

\begin{document}
\maketitle

\begin{abstract}
Deep neural networks (DNNs) have achieved substantial predictive performance in various speech processing tasks. Particularly, it has been shown that a monaural speech separation task can be successfully solved with a DNN-based method called deep clustering (DC), which uses a DNN to describe the process of assigning a continuous vector to each time-frequency (TF) bin and measure how likely each pair of TF bins is to be dominated by the same speaker. In DC, the DNN is trained so that the embedding vectors for the TF bins dominated by the same speaker are forced to get close to each other. 
One concern regarding DC is that the embedding process described by a DNN has a black-box structure, which is usually very hard to interpret. The potential weakness owing to the non-interpretable black-box structure is that it lacks the flexibility of addressing the mismatch between training and test conditions (caused by reverberation, for instance). To overcome this limitation, in this paper, we propose the concept of explainable deep clustering (X-DC), whose network architecture can be interpreted as a process of fitting learnable spectrogram templates to an input spectrogram followed by Wiener filtering. During training, the elements of the spectrogram templates and their activations are constrained to be non-negative, which facilitates the sparsity of their values and thus improves interpretability. The main advantage of this framework is that it naturally allows us to incorporate a model adaptation mechanism into the network thanks to its physically interpretable structure. We experimentally show that the proposed X-DC enables us to visualize and understand the clues for the model to determine the embedding vectors while achieving speech separation performance comparable to that of the original DC models. 

\end{abstract}

%================================================================================

\section{Introduction}

Time-frequency (TF) masking is widely used for monaural audio source separation. Given a monaural mixture speech signal, the goal is to estimate all the TF bins dominated by the same speaker, after applying the short-time Fourier transform (STFT) to the mixture signal. For TF mask prediction, there are two major approaches. 

One is a \textit{generative approach}, where the goal is to estimate the underlying parameters in a generative model describing the process by which a mixture signal has been generated. Typical examples of this approach include methods based on non-negative matrix factorization (NMF) \cite{Smaragdis2004nmf, Cichocki2006, Schmidt2006nmf, Virtanen2007, Fevotte2009} or non-negative matrix factor deconvolution (NMFD) \cite{Smaragdis2004nmfd, Schmidt2006nmfd, Smaragdis2007, Kameoka2009}. The idea behind these methods is to approximate the magnitude spectrogram of a mixture signal as a weighted sum of spectrum templates or spectrogram templates, pretrained using the speech samples of each speaker. Once the magnituide spectrogram of each speaker has been estimated from a mixture signal, we can use it to construct a Wiener mask that enhances the spectrogram of a particular speaker. The main advantage of these methods is that they easily allow us to incorporate a model adaptation mechanism into the generative process of observed signals, which makes it possible to compensate for the mismatch between the acoustic conditions at training and test time, caused, for example, by reverberation and unseen noise. However, one downside is that the separation performance can be limited due to the inconsistency between the training and test objectives. Namely, the process of training the spectrogram templates does not ensure that the separated signal at test time will be optimal. Another limitation is that, in principle, these methods can only work in speaker-dependent scenarios. 

The other is a \textit{discriminative approach}, where the goal is to directly learn to predict TF masks or TF embeddings from a given mixture signal. Methods based on the discriminative approach can be roughly categorized into class-based and partition-based methods. The class-based methods are designed to predict the most dominant speaker at each TF bin \cite{Han2011, Xu2014}, whereas the partition-based methods are designed to learn the embedding at each TF bin that can be used to measure how likely each pair of TF bins is to be dominated by the same speaker \cite{Bach2006, Hershey2016, Li2018}. 
A partition-based method incorporating a deep neural network (DNN) called \textit{deep clustering} (DC) \cite{Hershey2016, Li2018} has attracted a lot of attention in recent years owing to its several attractive features. The main idea of DC is to learn an embedding for each TF bin of an input spectrogram so that a TF mask for each speaker can be constructed by clustering the embedding vectors. 
Instead of using a DNN to directly predict the speaker label for each TF point, DC uses a DNN to describe the embedding process and trains it so that the embedding vectors of each pair of TF bins dominated by the same speaker get close to each other. 
This allows the DNN to produce an output that is invariant to the number and permutations of the speaker labels. This property, called the permutation-free property, is particularly favorable in that it allows the model to handle even speaker-independent speech separation (i.e., separation of unseen speakers). Several attempts have been made to develop models incorporating this property since DC was proposed. These include Permutation Invariant Training (PIT) \cite{Yu2017} and Deep Attractor Net (DANet) \cite{Chen2017}. 
The model proposed in this paper is also designed to satisfy the permutation-free property. 

While the above DNN-based discriminative methods are often reported to perform better than the generative counterparts, the decision-making process of DNNs is usually difficult to interpret. For instance, in DC \cite{Hershey2016, Li2018}, the trained DNN does not provide any reasons as to why it has come up with the specific value of each embedding vector according to an input spectrogram. One potential weakness in DC owing to the non-interpretable black-box structure is that it is not flexible enough to address the mismatch between training and test conditions (caused by reverberation, for instance). If we could somehow describe the DNN based on the generative process of observed signals, we would be able to incorporate a model adaptation mechanism into its process, according to which we could compensate for the mismatch at test time. 

In this paper, to address the above concerns, we propose an extension of DC called \textit{explainable deep clustering} (\textbf{X-DC}), where the embedding process is described by a DNN with a physically interpretable structure. Specifically, we design the network architecture so that it can be interpreted as a process of fitting learnable short-segment spectrogram templates to an input spectrogram followed by Wiener filtering. Namely, in our model, the square root Wiener mask vector, in which each element indicates the square root of the probability of a particular speaker being present in a mixture, is treated as the embedding vector. Thus, the dimension of the embedding space can be seen as the maximum potential number of speakers. Notice that a weighted sum of spectrogram templates shifted at all different positions can be described by an NMFD model \cite{Smaragdis2004nmfd, Schmidt2006nmfd, Smaragdis2007, Kameoka2009} (i.e., a convolutive mixture of spectrogram templates). Using this fact, we design the X-DC model so that the network mainly consists of the ``NMFD" part and the ``Wiener mask construction" part, where the former produces a set of the non-negative activation matrices in the NMFD models for all possible sources and the latter is designed to produce Wiener masks constructed according to the NMFD models, each of which can be described as the output of a convolution layer with non-negative kernels. Since we would like the intermediate layer outputs of this network to correspond to the estimates of the spectrograms of underlying sources, we include a training loss to encourage the sum of them to become close to the input spectrogram, in addition to the original training loss for DC measured using the embedding vectors. This allows the Wiener mask construction process to be structure-aware in the sense that Wiener masks are constructed explicitly according to the spectrogram estimates. Another potential advantage of X-DC is that it easily allows us to incorporate a model adaptation mechanism into the network thanks to its physically interpretable structure. For example, we may be able to make it handle unseen reverberation by inserting a non-negative convolution layer corresponding to the reverberation process \cite{Kameoka2009} into the network and optimize it at test time so that we can jointly perform dereverberation and Wiener-mask inference. In Section \ref{sec:X_DC}, we give a detailed explanation of the proposed X-DC model. 

In Section \ref{sec:experiment}, we experimentally show that the proposed X-DC enables us to visualize and understand the clues for the model to determine the embedding vectors while achieving speech separation performance comparable to the original DC. Furthermore, the experimental results show that the trained spectrogram templates obtained with X-DC captured the harmonic structure of speech fairly well, even though the proposed X-DC model was not explicitly implemented to use harmonic structure information. 

\paragraph{Related work (1) NMFD.} The original NMFD \cite{Smaragdis2004nmfd} is an extension of the idea of using the NMF model to express audio spectrograms \cite{Smaragdis2004nmf}. NMF is a method for factorizing a given non-negative matrix $A = (a_{f, n})_{1 \leq f \leq F, 1 \leq n \leq N}$ (e.g., a magnitude spectrogram in the speech separation task) \footnote{We use the term ``non-negative matrix'' to mean the matrix all of whose entries are non-negative.} into the product of two non-negative matrices $W = (w_{f, j})_{1 \leq f \leq F, 1 \leq j \leq J}$ and $H = (h_{j, n})_{1 \leq j \leq J, 1 \leq n \leq N}$: $A \approx WH$, where we generally set $J < \min \{ F, N \}$. Namely, each $(f, n)$th entry of $A$ is approximated by
\begin{align}
\label{eq:nmf_original}
A_{f, n} \approx \sum_{j=1}^J w_{f, j} h_{j, n}. 
\end{align}
The two matrices $W$ and $H$ can be found by iteratively decreasing the difference between $A$ and $WH$ (see \cite{Paatero1997, Lee1999, Lee2001, Berry2007} for specific algorithms) until convergence, starting from random initial values. Like NMF, NMFD finds a decomposition of $A$, but is based on a convolutive mixture model instead of the instantaneous mixture model (\ref{eq:nmf_original}). Namely, each $(f, n)$th entry of the observed matrix $A$ is modeled as a sum of every shifted version of $\{ W_j \}$ scaled by $\{ \bm{h}_j \}$: 
\begin{align}
\label{eq:nmfd_original}
A_{fn} \approx \sum_{j=1}^J \sum_{m=1}^M w_{j, f, m} h_{j, n - m + 1}, 
\end{align}
where $W_j = (w_{j, f, m})_{1 \leq f \leq F, 1 \leq m \leq M}$ and $\bm{h}_j = (h_{j, n})_{1 \leq n \leq N}$, respectively, represent the $j$th template matrix and its activation vector. As can be seen from (\ref{eq:nmf_original}) and (\ref{eq:nmfd_original}), NMFD reduces to NMF when $M=1$. Owing to the time index $m$ in $W$, NMFD allows each template to express the local spectro-temporal pattern underlying an observed spectrogram. 
The proposed X-DC model incorporates the expression (\ref{eq:nmfd_original}) into the last layer of a neural network so that the model output can be interpreted as a convolutive mixture of spectrogram templates. There are differences between the original NMFD and the proposed X-DC in model formulation and loss function. 
First, while NMFD treats each element of $W$ and $H$ as a free variable to estimate, X-DC treats $W$ as the kernel values of a convolution layer and $H$ as the output of the previous layer in a DNN. Second, while source separation using NMFD is categorized as a generative approach, where the objective is to approximate an observed mixture spectrogram as described above, X-DC is a discriminative one, where the objective is to find the best partition of the TF embeddings, es explained in Section \ref{sec:X_DC}.

\paragraph{Related work (2) DeepNMF.} The idea of using a single DNN to describe the processes of fitting the NMF model to an input spectrogram, constructing Wiener masks, and generating the outputs of the Wiener masks has already been adopted in a concept called DeepNMF \cite{Hershey2014, Roux2015, Wisdom2017}. In DeepNMF, each intermediate layer of the DNN is interpreted as a single iteration of the multiplicative update algorithm for NMF and the final layer involves a Wiener filtering-like operation. 
Specifically, let $W^k \equiv \begin{bmatrix}
W^{(k, 1)} & \cdots & W^{(k, I)}
\end{bmatrix}$ and $H^k \equiv \begin{bmatrix}
\left( H^{(k, 1)} \right)^{\mathsf{T}} & \cdots & \left( H^{(k, I)} \right)^{\mathsf{T}}
\end{bmatrix}^{\mathsf{T}}$, respectively, be the parameter of the $k$th layer and the input to the same layer, which can also be interpreted as the NMF parameters in the $k$th iteration ($k = 1, \dots, K$). Here, $W^{(k, i)}$ and $H^{(k, i)}$ are the parameters corresponding to the $i$th speaker ($i = 1, \dots, I$). In NMF-based speech separation, the spectrogram of each $i$th speaker given the finally estimated parameters $W^K$ and $H^K$ is typically obtained as
\begin{align}
\label{eq:deepNMF_WF}
S^{(K, i)} = \frac{W^{(K, i)} H^{(K, i)}}{\sum_{i=1}^I W^{(K, i)} H^{(K, i)}} \circ A, 
\end{align}
where $\circ$ represents element-wise multiplication. The above Wiener filtering-like operation (\ref{eq:deepNMF_WF}) can be incorporated in the final layer of the DNN, and thus the entire DNN represents a function that maps the initial NMF parameter $H^0$ to the separated spectrograms $\{ S^{(K, i)} \}_{i = 1, \dots, I}$. The DNN is trained so that it can infer the NMF parameters $W^K$ and $H^K$ in the second last layer, and the spectrogram of each speaker (as the output of the DNN) from an observed mixture spectrogram. Our X-DC differs from DeepNMF in several aspects. First, X-DC uses the NMFD model instead of the regular NMF model as the building block for the architecture design. Second, the shallower part of the X-DC architecture can be designed freely and does not need to be expressed in the form of the multiplicative update rule of NMFD. Finally, the most essential difference lies in the permutation-invariance property: Since DeepNMF employs a training objective given by the difference between the model outputs and the spectrograms of target speech, the model training depends on the permutations of the source order and is thus \textit{not} permutation-invariant. This implies that DeepNMF can only handle speaker-dependent speech separation. By contrast, X-DC differs from DeepNMF in that our model uses a training objective that is permutation-invariant as with DC, thus allowing it to handle speaker-independent separation. 

In the following sections, we first describe the original DC \cite{Hershey2016, Li2018} in Section \ref{sec:DC}. Then, we explain how to construct the proposed X-DC model in Section \ref{sec:X_DC}. In Section \ref{sec:experiment}, we describe the experiment we conducted to compare the speech separation performance of the proposed X-DC and the conventional DC models, and to check the extracted spectrogram templates and their activations in the proposed X-DC. We conclude this paper in Section \ref{sec:conclusion}. 

%================================================================================

\section{Original Deep Clustering}
\label{sec:DC}

Before describing our proposed model X-DC, we briefly review the original concept of DC \cite{Hershey2016, Li2018}. Let $X = (X_{f, n})_{1 \leq f \leq F, 1 \leq n \leq N} \in \mathbb{R}^{F \times N}$ be the spectrogram of a mixture signal of $I$ sources. We denote the vectorized spectrogram as $\bm{x} = \mathrm{vec} (X) = \left[ x_1, \dots, x_k, \dots, x_K \right]^{\mathsf{T}} \in \mathbb{R}^K$, where the index $k = F \times (n - 1) + f \in \{ 1, \dots, K \}$ corresponds to the $(f, n)$th TF point of the original spectrogram and $K = F \times N$. From the input spectrogram, a DNN is used to map each $k$th TF bin into a $D$-dimensional unit-norm embedding vector $\bm{v}_k = \left[ v_{k, 1}, \dots, v_{k, D} \right]^{\mathsf{T}}$. Based on these embedding vectors, we define a matrix $V = \left[ \bm{v}_1, \dots, \bm{v}_{K} \right]^{\mathsf{T}} \in \mathbb{R}^{K\times D}$. 

The main idea of DC is to train a nonlinear mapping $V = \varphi_{\theta} (\bm{x})$ from input spectrogram $\bm{x}$ to the embedding vectors $V$ with a set of model parameters $\theta$ so that the embedding vectors of the TF bins dominated by the same source be close to each other. We can use either a recurrent neural network (RNN) \cite{Hershey2016} or a convolutional neural network (CNN) \cite{Li2018} to implement $\varphi_{\theta}$ to let each element of $\varphi_{\theta} (\bm{x})$ be determined according to the entire range of $\bm{x}$. 
Let $\bm{y}_k = (y_{k, i})_{1 \leq i \leq I} \in \{0, 1\}^I$ be a one-hot vector that indicates which speaker is dominant at the $k$th TF point in input spectrogram $\bm{x}$, where $I$ is the maximum potential number of speakers. That is, $y_{k, i} = 1$ if the $i$th speaker is dominant at the $k$th TF point, and $y_{k, i} = 0$ otherwise. Based on these vectors, we define a matrix $Y = \left[ \bm{y}_1, \dots, \bm{y}_K \right]^{\mathsf{T}} \in \{0, 1\}^{K \times I}$. 

From these definitions, matrices $V V^{\mathsf{T}} \in \mathbb{R}^{K \times K}$ and $Y Y^{\mathsf{T}} \in \{ 0, 1 \}^{K \times K}$ represent the estimated and true affinity matrices of all the TF points. 
The $(k, k')$th element of matrix $V V^{\mathsf{T}}$ represents the similarity between the embedding vectors at the $k$th and $k'$th TF bins, whereas that of matrix $Y Y^{\mathsf{T}}$ is one if the $k$th and $k'$th TF bins belong to the same speaker, and is zero otherwise. At training time, we aim to let $V V^{\mathsf{T}}$ get as close as possible to $Y Y^{\mathsf{T}}$. 
Thus, the loss function of DC for a set of model parameters $\theta$ is given by
\begin{align}
\tilde{\mathcal{J}} (\theta) 
&= \| V V^{\mathsf{T}} - Y Y^{\mathsf{T}} \|_{\mathrm{F}}^2 \nonumber \\
&= \| V^{\mathsf{T}} V\|_{\mathrm{F}}^2 - 2\| V^{\mathsf{T}} Y \|_{\mathrm{F}}^2 + \| Y^{\mathsf{T}} Y \|_{\mathrm{F}}^2. 
\label{eq:DC}
\end{align}
Here, $\| \cdot \|_{\mathrm{F}}^2$ is the Frobenius norm of a matrix. It must be noted that we can easily confirm that $Y Y^{\mathsf{T}}$ is invariant under the permutations of the speaker order and thus leads to the permutation-free property of DC. 

At test time, given input spectrogram $\bm{x}$, we first compute $V = \varphi_{\theta} (\bm{x})$ and then perform clustering (e.g., k-means algorithm) on the rows of $V$, that is, the embedding vectors $\{ \bm{v}_k \}$ of all the TF points. Each obtained cluster consists of the TF points with similar embedding vectors, according to which we can determine the TF mask for a particular speaker. 

As described above, an important advantage of DC is that it is invariant under the permutations of speaker labels. Namely, speaker labels do not need to be consistent over different utterances in the training data. Thanks to this permutation-free property, the DC model has been shown to generalize well over unseen speakers and languages, presumably by capturing speaker- and language-independent patterns (such as the harmonic structure and temporal/spectral continuity) underlying the spectrogram of a single stream of speech. 

%================================================================================

\section{X-DC: Explainable Deep Clustering}
\label{sec:X_DC}

\begin{figure}
\centering
 \begin{minipage}{\hsize}
 \centering
 \includegraphics[width=0.9\hsize]{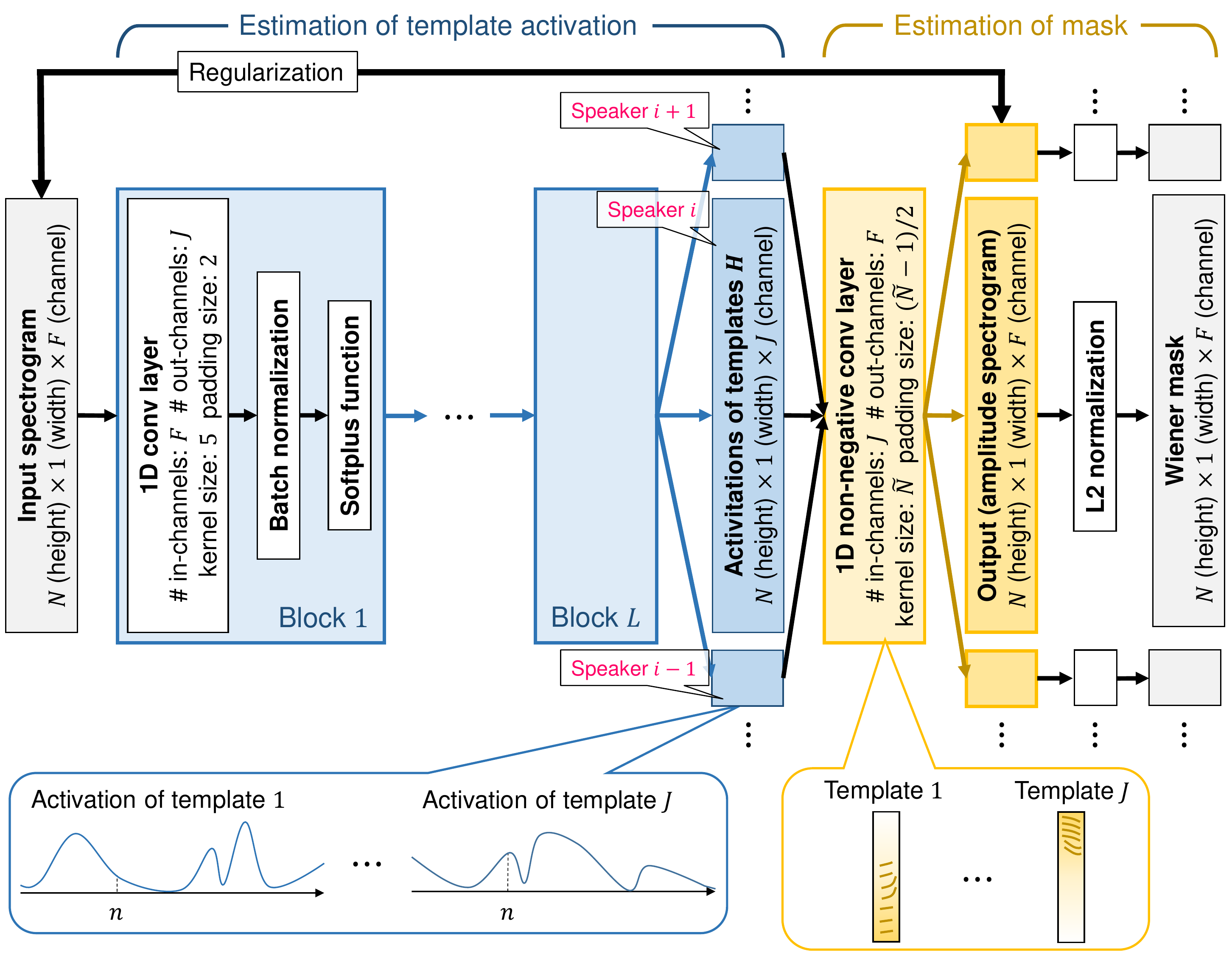}\\
 \includegraphics[width=0.75\hsize]{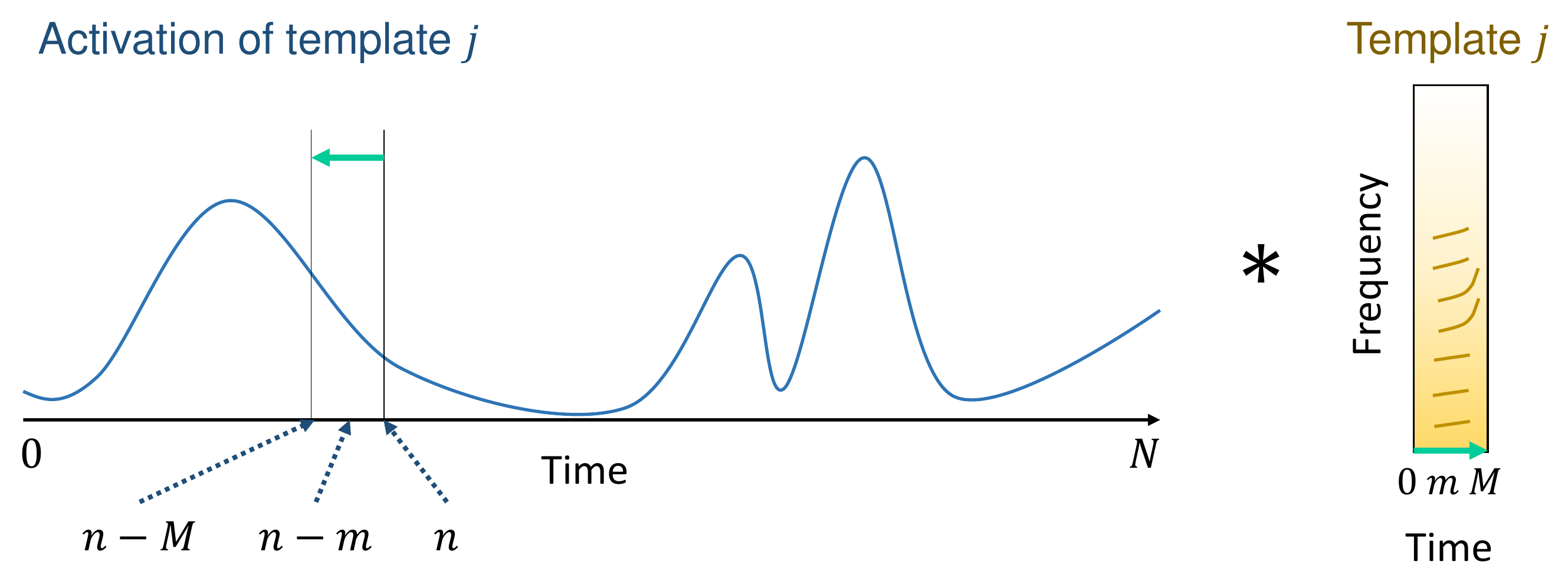}
 \end{minipage}
\caption{Neural network architecture of the proposed X-DC. An amplitude spectrogram is input to the proposed network, and first its embedding vectors are extracted, which can be interpreted as activation vectors of the spectrogram templates later. An arbitrary network architecture can be used as such a shallower part of the network for extracting embedding vectors only if the output activations are constrained to be non-negative. Then, the extracted embeddings are input to the deeper part of the network, where a Wiener mask for each speaker is estimated based on the embeddings. This mask construction part consists of a one-dimensional convolutional layer with non-negative convolutional kernels, which can be interpreted as learnable spectrogram templates. Finally, the output of the convolutional layer is normalized so that the output of the entire network can be viewed as a square root Wiener mask for each speaker. To make the output before normalization identifiable and interpretable as an amplitude spectrogram, we add a regularization term to the loss function that penalizes the discrepancy between the input spectrogram and the output before normalization. The bottom part of the figure shows how we can interpret the estimation process of the proposed X-DC. The estimation is performed by the convolution of the trained spectrogram templates and their activations. }
\label{fig:nn}
\end{figure}

In this section, we propose an extension of DC named \textit{explainable deep clustering} (X-DC) and describe its specific formulation in detail. Figure \ref{fig:nn} shows the entire network architecture of the proposed X-DC. The proposed X-DC inherits the main advantages of the conventional DC described in Section \ref{sec:DC}, while explicitly incorporating the process of fitting learnable spectrogram templates to input spectrograms. 

By taking account of the unit norm constraint, we start by treating a square root Wiener mask vector as the TF embedding vector $\bm{v}_k$. Namely, each element of vector $\bm{v}_k$ is given by the square root of the Wiener mask for a different speaker. 
We believe that this is reasonable since it is easy to see that clustering these vectors directly corresponds to finding all the TF bins that are dominated by the same speaker. To define $\bm{v}_k$, we start out by defining $\tilde{v}^{(i)}_{f, n}$ as
\begin{align}
\label{eq:v}
&\tilde{v}^{(i)}_{f, n} = \frac{\tilde{h}^{(i)}_{f, n}}{\sqrt{\sum_{i'} \left( \tilde{h}^{(i')}_{f, n} \right)^2} + \epsilon}, \nonumber \\
&f = 1, \dots, F,\ \ n = 1, \dots, N,\ \ i = 1, \dots, I, 
\end{align}
where $\tilde{v}^{(i)}_{f, n}$ and $\tilde{h}^{(i)}_{f, n}$ denote the outputs of the last and second last layers at the $i$th channel corresponding to the $(f, n)$th TF point, respectively. To avoid the division by zero, notice that we have added a small constant $\epsilon$ to the denominator of each mask value $\tilde{v}^{(i)}_{f, n}$ in (\ref{eq:v}). In the following experiment, we set this constant at $\epsilon = 10^{-5}$. The vector $\bm{v}_k = [v_{k, 1}, \dots, v_{k, I}]$ is finally obtained by arranging $\tilde{v}^{(1)}_{f, n}, \dots, \tilde{v}^{(I)}_{f, n}$ into a vector such that $v_{F (n - 1) + f, i} = \tilde{v}^{(i)}_{f, n}$. By further defining $V = [\bm{v}_1, \dots, \bm{v}_K]^{\mathsf{T}}$, we can use (\ref{eq:DC}) as is as our training objective. 

Since $\tilde{v}^{(i)}_{f, n}$ is treated as a square root Wiener mask for the $i$th speaker, $\tilde{h}^{(i)}_{f, n}$ must correspond to the magnitude of the $i$th speaker at the $(f, n)$th TF bin. In this view, we would like to ensure that the sum of $\tilde{h}^{(i)}_{f, n}$ over $i = 1, \dots, I$ is consistent with the input magnitude spectrogram $X_{f, n}$. To this end, we consider including the loss
\begin{align}
\mathcal{R} (\theta) 
&= \frac{\lambda}{4K} \| X - \sum_i \tilde{H}^{(i)} \|_{\mathrm{F}}^2, 
\label{eq:regularization}
\end{align}
in our training objective, where $\tilde{H}^{(i)} \equiv (\tilde{h}^{(i)}_{f, n})_{1 \leq f \leq F, 1 \leq n \leq N}$, and $\lambda$ is a regularization hyperparameter. It is important to note that inclusion of this regularization term does not violate the permutation-invariance property since this term is also invariant under the permutations of speaker indices $i = 1, \dots, I$. 

Since the spectrograms of speech have certain structures that are common across different speakers and languages (such as the harmonic/formant structure and spectral/temporal continuity), it would be natural to assume that the spectrogram of each speaker can be expressed as a superposition of local spectrogram templates drawn from a common dictionary. Thus, one way to express the magnitude spectrogram $\tilde{h}^{(i)}_{f, n}$ would be to assume an NMFD-like model \cite{Smaragdis2004nmfd} such that
\begin{align}
\tilde{h}^{(i)}_{f, n} &= \sum_{j = 1}^J \sum_{m = 1}^M w_{j, f, m} h^{(i)}_{j, n-m+1} \nonumber \\
&= \sum_{j = 1}^J \sum_{m = n - M}^{n - 1} w_{j, f, n-m} h^{(i)}_{j, m+1}. 
\label{eq:conv_xdc}
\end{align}
If we use $H = (h_{q, n})_{1 \leq q \leq Q, 1 \leq n \leq N} \in \mathbb{R}^{Q \times N}$ to denote the output of the previous layer with channel number $Q$ and length $N$, (\ref{eq:conv_xdc}) can be described as a parallel convolution layer. Namely, it can be seen as a layer consisting of vertically splitting $H$ into $I$ sub-arrays $H^{(1)}, \dots, H^{(I)} \in \mathbb{R}^{J \times N}$ with the same shape (i.e., $Q = IJ$), applying a regular 1D convolution to these sub-arrays, treated as a virtual mini-batch, and producing $\tilde{H}^{(1)}, \dots, \tilde{H}^{(I)} \in \mathbb{R}^{F \times N}$ in parallel. Here, $J$ denotes the number of the spectrogram templates and $M$ denotes the length of each template. 
Simply put, the magnitude spectrogram $\tilde{h}^{(i)}_{f, n}$ of the $i$th speaker is assumed to be given as a convolutive mixture of the $J$ spectrogram templates, as shown at the bottom of Figure \ref{fig:nn}. 

To interpret the convolution kernel $W$ as a set of magnitude spectrogram templates and the layer input $H$ as a set of the corresponding temporal activity patterns, we impose a non-negative constraint on each element of both $W$ and $H$, assuming the additivity of magnitude spectrograms (though this holds only approximately). This non-negative constraint is expected to induce the sparsity of $W$ and $H$ so that the trained $W$ can be interpreted as non-negative principal parts that frequently appear in $X$. We can implement the non-negative constraint on the activations $H$ by including an activation function that is ensured to produce non-negative values only (e.g., a softplus and rectified linear unit function) in the third last layer. As for the convolution kernel $W$, we replace each element with $w_{j, f, n} =  \max \{ 0, \tilde{w}_{j, f, n} \}$ and treat the unconstrained variable $\tilde{W}=(\tilde{w}_{j, f, n})_{j, f, n}$ as the parameter to train instead. 

The remaining layers can be designed arbitrarily under two requirements: the input of the network must be a spectrogram (2D array) with $F \times N$ TF bins, and the output of the third last layer $H$ must be a 3D array with shape $J \times N \times I$ with non-negative elements. The part of the entire network from the input layer to the third last layer can be thought of as an NMFD part, which is responsible for predicting the template activity patterns from the input spectorgram whereas the remaining part can be seen as a Wiener mask construction part, which constructs a square root Wiener mask in accordance with the spectrogram templates and the corresponding activity patterns. We train the entire network function $\varphi_{\theta}$ of the proposed X-DC model with the following loss function: 
\begin{align}
\mathcal{J} (\theta) = \tilde{\mathcal{J}} (\theta) + \mathcal{R} (\theta). 
\label{eq:XDC}
\end{align}

At test time, we can construct a TF mask directly from $\tilde{v}^{(i)}_{f, n}$ given an input spectrogram $X$. The TF mask obtained in this way is expected to produce reasonably natural and smooth spectrograms since it respects the constraint given by the spectrogram estimate of each source. 

%%%
%%%

%================================================================================
\section{Experiments}
\label{sec:experiment}

To show the effectiveness of the proposed X-DC model, we conducted a speech separation experiment using the CMU Arctic speech databases \cite{arctic2004} and L2-ARCTIC \cite{arctic2018} to compare its performance with the conventional DC models \cite{Hershey2016, Li2018} and to check the obtained spectrogram templates and their activations. 

%---
\begin{table*}[t]
\centering
\caption{Experimental settings of the speakers in the training, validation, and test data sets. ``Known'' shows that the pair of speakers in the test data set has also been in the training data set, and ``unknown'' shows otherwise. M and F, respectively, stand for male and female speakers.}\vspace{0.02\hsize}
\begin{tabular}{|c||p{0.25\hsize}|c|c|c|c|} \hline 
\rowcolor[rgb]{0.93, 0.93, 0.93} \multicolumn{1}{|c||}{\textbf{Setting \#}} & \multicolumn{1}{c|}{\textbf{Training data}} & \multicolumn{1}{c|}{\textbf{Validation data}} & \multicolumn{3}{c|}{\textbf{Test data}} \\ \hline \hline
$1$ & \multirow{6}{*}{\parbox{\hsize}{\textbf{$2$ speakers} (bdl, clb)}} & \multirow{18}{*}{bdl, clb} & \multirow{3}{*}{Known} & M/F & bdl, clb \\
$2$ & & & & M/M & - \\
$3$ & & & & F/F & - \\ \cline{4-6}
$4$ & & & \multirow{3}{*}{Unknown} & M/F & rms, slt \\
$5$ & & & & M/M & rms, aba \\
$6$ & & & & F/F & slt, zhaa \\ \cline{2-2} \cline{4-6}
$7$ & \multirow{6}{*}{\parbox{\hsize}{\textbf{$4$ speakers} (bdl, clb, rms, slt)}} & & \multirow{3}{*}{Known} & M/F & bdl, clb \\
$8$ & & & & M/M & bdl, rms \\
$9$ & & & & F/F & clb, slt \\ \cline{4-6}
$10$ & & & \multirow{3}{*}{Unknown} & M/F & aba, zhaa \\
$11$ & & & & M/M & aba, bwc \\
$12$ & & & & F/F & zhaa, lxc \\ \cline{2-2} \cline{4-6}
$13$ & \multirow{6}{*}{\parbox{\hsize}{\textbf{$8$ speakers} (bdl, clb, rms, slt, aba, zhaa, bwc, lxc)}} & & \multirow{3}{*}{Known} & M/F & bdl, clb \\
$14$ & & & & M/M & bdl, rms \\
$15$ & & & & F/F & clb, slt \\ \cline{4-6}
$16$ & & & \multirow{3}{*}{Unknown} & M/F & asi, svbi \\
$17$ & & & & M/M & asi, hkk \\
$18$ & & & & F/F & svbi, hjk \\ \hline 
\end{tabular}
\label{tb:setting_speaker}
\end{table*} 
%---

Table \ref{tb:setting_speaker} shows the experimental settings of the speakers in the training, validation, and test data sets. As a training data set, we used $N^{\mathrm{sample}}$ sets of a mixed speech signal of two, four, or eight speakers and the corresponding true affinity matrix $Y Y^{\mathsf{T}}$, where $N^{\mathrm{sample}} \equiv \max \{ N^{(1)}, N^{(2)} \}$ and $N^{(i)}$ is the number of utterances in the training data set of the $i$th speaker. In each epoch, a pair of speakers, $1$ and $2$, were randomly selected from the $2$, $4$, or $8$ speakers shown in Table \ref{tb:setting_speaker} from the uniform distribution. 
To make the mixed speech signal, we first applied the STFT based on the Hanning window to the speech signal of each speaker, and obtained the spectrogram $\tilde{X}^{(i)}$ of the $i$th speaker for all $i = 1, \dots, I$ ($I = 2$ in this case) by multiplying random weights that were generated from the uniform distribution on $[0, 1)$. From these spectrograms, we computed the scaled amplitude spectrogram $X$ of a mixed signal by
\begin{align}
&X = (X_{f, n})_{1 \leq f \leq F, 1 \leq n \leq N}, \nonumber \\
&X_{f, n} = | \tilde{X}_{f, n} | / \max_{f, n} |\tilde{X}_{f, n}|, 
\end{align}
where $\tilde{X} = \sum_i \tilde{X}^{(i)}$ is the complex spectrogram of the mixed signal. With regard to the input to the conventional DC models, we used the log spectrogram $X_{f, n} = 10 \log_{10} (|\tilde{X}_{f, n}| + 10^{-16})$, as in \cite{Li2018}. We considered the $(f, n)$th TF point to be silent (i.e., it is not assigned to any speaker) if and only if the following inequality holds: $20 \log_{10} (|\tilde{X}_{f, n}| / \max_{f, n} (|\tilde{X}_{f, n}|) +10^{-16}) < -40$. 
Finally, to make the true affinity matrix, we defined the one-hot vectors of the speaker labels by $Y = \left[ \bm{y}_1, \dots, \bm{y}_K \right]^{\mathsf{T}}$, where $\bm{y}_k = (y_{k, i})_{1 \leq i \leq I} \in \{0, 1\}^I$, $k = F \times (n - 1) + f \in \{ 1, \dots, K \}$, and 
\begin{eqnarray}
y_{k, i} \equiv \begin{cases}
1 & \mathrm{if}\ i = \mathrm{argmax}_{i'} |\tilde{X}^{(i')}_{f, n}|, \\
0 & \mathrm{if\ otherwise}, 
\end{cases}
\ \mathrm{for\ all}\ (f, n). 
\end{eqnarray}
Based on such speaker labels $Y$, we obtained an affinity matrix $Y Y^{\mathsf{T}}$ of all the pairs of TF points. 

As for a test data set, we used $66$ sets of mixed speech signals of the two speakers shown in Table \ref{tb:setting_speaker}, and obtained the sets of input and output data based on the same procedure as the training data set. The only difference is that we multiplied the spectrogram of each speaker by the deterministic weight of one for the test data set, not by a random weight. 

Using the above test data set, we compared the speech separation performance of the conventional DC models \cite{Hershey2016, Li2018} and the proposed X-DC model, all of which were trained with the above training data set. As performance measures, we used the source to distortion ratio (SDR), source to interference ratio (SIR), and source to artifact ratio (SAR) \cite{Vincent2006}. 

The detailed numerical settings are as follows: 
\begin{itemize}
\item For the conventional DC model of a BLSTM network \cite{Hershey2016}, we set the dimension of the embedding space at $D = 20$, the number of hidden cells in each BLSTM layers at $600$, the number of the BLSTM layers at $3$, the learning rate at $10^{-5}$, and the number of epochs for training at $T = 1100$. Under these settings, the number of learnable parameters in the conventional DC model of a BLSTM network is $23,880,160$.
\item For the conventional DC model of a gated convolutional network \cite{Li2018}, 
  \begin{itemize}
  \item When there are two speakers in the training data set, we set the dimension of the embedding space at $D = 90$, the number of output channels in the middle convolutional layers at $C = 14$, the number of the middle convolutional layers at $L^{\mathrm{conv}} = 3$ (layers) $\times N^{\mathrm{block}}$ (blocks) with $N^{\mathrm{block}} = 30$, the learning rate at $\eta = 5 \times 10^{-4}$, and the number of epochs for training at $T = 700$. Under these settings, the number of learnable parameters in the conventional DC model of a gated convolutional network is $1,181,802$.
  \item When there are four speakers in the training data set, we set $D = 10$, $C = 18$, $L^{\mathrm{conv}} = 3$ (layers) $\times N^{\mathrm{block}}$ (blocks) with $N^{\mathrm{block}} = 20$, $\eta = 5 \times 10^{-3}$, and $T = 1,000$. Under these settings, the number of learnable parameters is $1,143,154$.
  \item When there are eight speakers in the training data set, we set $D = 60$, $C = 22$, $L^{\mathrm{conv}} = 3$ (layers) $\times N^{\mathrm{block}}$ (blocks) with $N^{\mathrm{block}} = 15$, $\eta = 5 \times 10^{-4}$, and $T = 800$. Under these settings, the number of learnable parameters is $1,169,316$.
  \end{itemize}
\item For the proposed X-DC model, we set the maximum potential number of speakers at $I = 2$, and used the following settings: 
  \begin{itemize}
  \item When there are two speakers in the training data set, we set the regularization hyperparameter at $\lambda = 10^{-3}$, the frame width of a convolutional kernel at $M = 15$, the number of spectrogram templates at $J = 40$, the number of output channels in the middle convolutional layers at $C = 241$, the number of layers of ``NMFD part'' at $L^{\mathrm{NMFD}} = 5$, the learning rate at $\eta = 10^{-2}$, and the number of epochs for training at $T = 700$. Under these settings, the number of learnable parameters in the proposed X-DC model is $1,201,787$.
  \item When there are four speakers in the training data set, we set $\lambda = 5 \times 10^{-2}$, $M = 15$, $J = 60$, $C = 185$, $L^{\mathrm{NMFD}} = 7$, $\eta = 5 \times 10^{-3}$, and $T = 1,100$. Under these settings, the number of learnable parameters is $1,203,915$.
  \item When there are eight speakers in the training data set, we set $\lambda = 5 \times 10^{-1}$, $M = 15$, $J = 120$, $C = 179$, $L^{\mathrm{NMFD}} = 6$, $\eta = 5 \times 10^{-3}$, and $T = 700$. Under these settings, the number of learnable parameters is $1,203,985$.
  \end{itemize}
\item As a common setting to all the models, we set the batch size at $4$, the sampling rate at $8$ khz, the window size of the STFT at $254$ points (which results in that the number of frequency bins is $F = 254/2 + 1 = 128$), the window shift at $127$ points, and the number of time bins of an input spectrogram at $N = 100$. As a training algorithm, we used Adam \cite{Kingma2015}. The number of utterances in the training data set of each speaker is as follows: $1,000$ (bdl, clb, rms, slt, zhaa, svbi), $999$ (lxc, asi, hkk, hjk), $998$ (bwc), and $997$ (aba).
\end{itemize}
The above hyperparameter settings (i.e., $T$ for the BLSTM-DC, $D$, $C$, $N^{\mathrm{block}}$, $\eta$, and $T$ for the conv-DC, and $J$, $C$, $L^{\mathrm{NMFD}}$, $\eta$, $T$, and $\lambda$ for the proposed X-DC) were chosen by the hold-out validation with $66$ validation data sets.

\begin{figure*}[p]
\centering
\includegraphics[width=0.95\hsize]{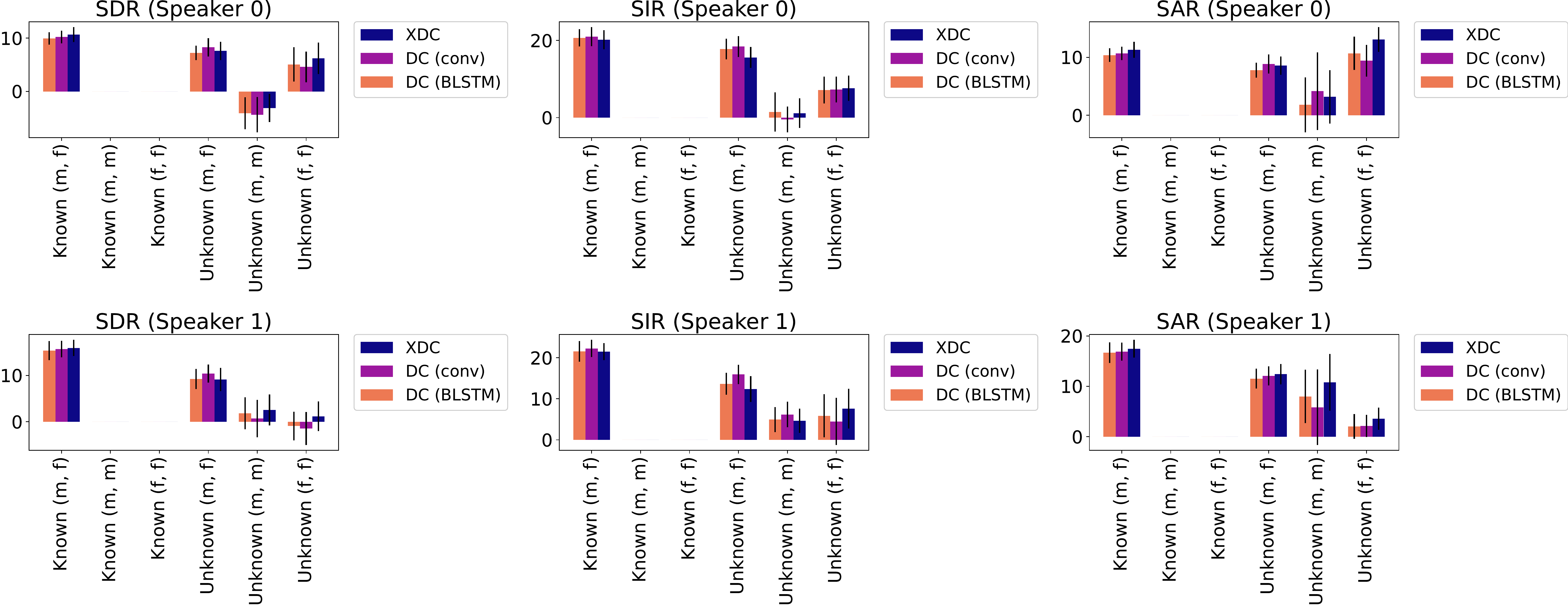}\vspace{-2mm}
\caption{Comparison of speech separation performance between the conventional DC models and proposed X-DC model (\textbf{$2$ speakers} in the training data set). Each bar shows the mean and standard deviation of the results for the test data set.}\vspace{5mm}
\label{fig:SDR2}
\includegraphics[width=0.95\hsize]{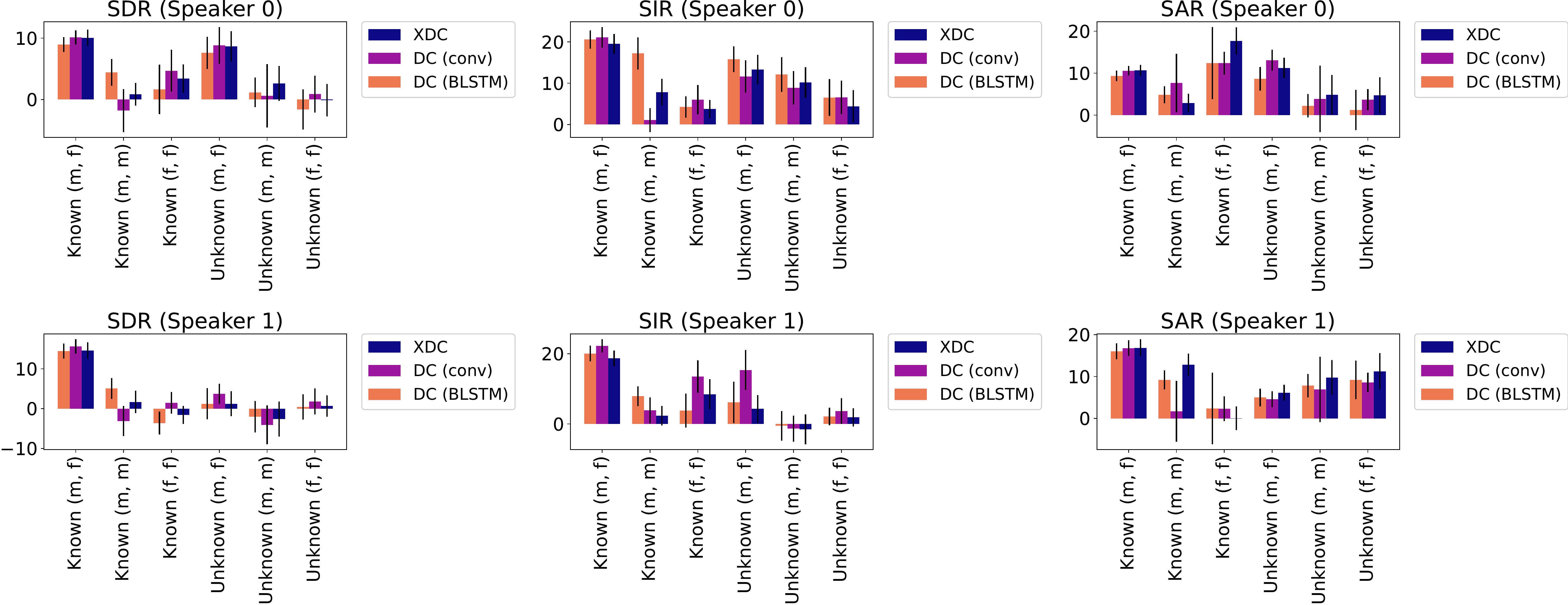}\vspace{-2mm}
\caption{Comparison of speech separation performance between the conventional DC models and proposed X-DC model (\textbf{$4$ speakers} in the training data set).}\vspace{5mm}
\label{fig:SDR4}
\includegraphics[width=0.95\hsize]{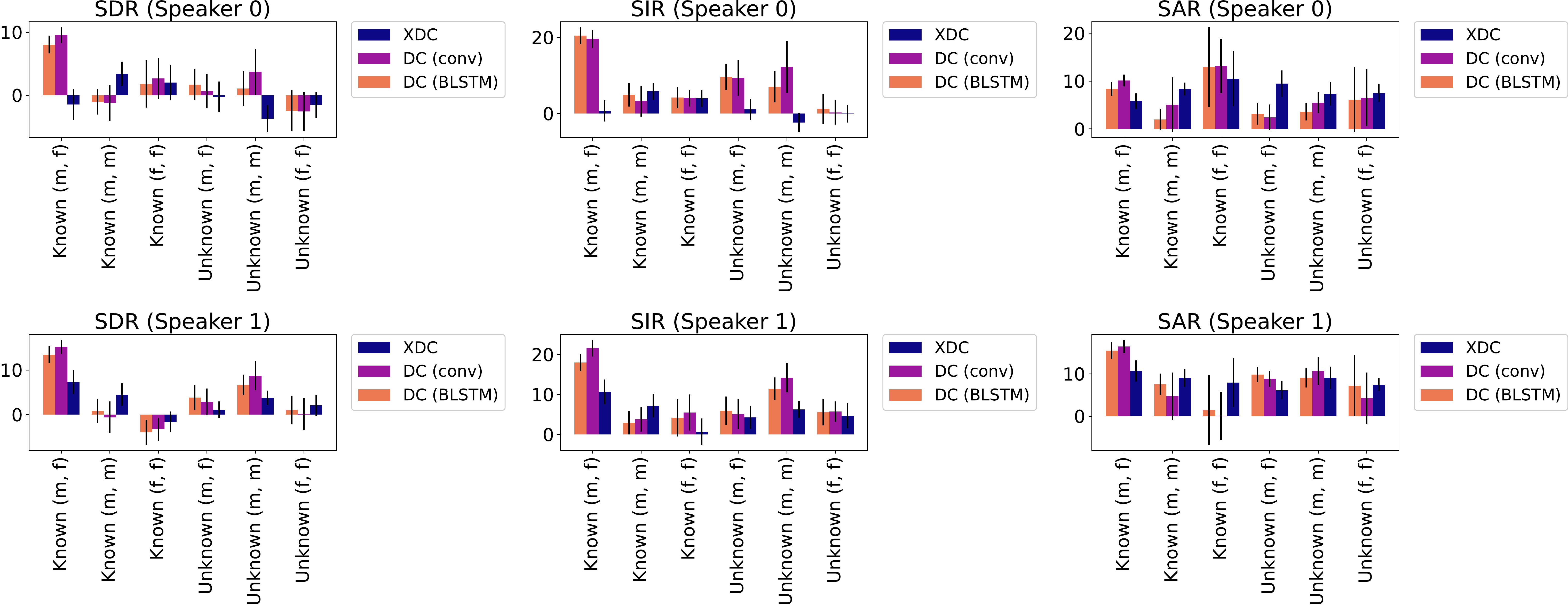}\vspace{-2mm}
\caption{Comparison of speech separation performance between the conventional DC models and proposed X-DC model (\textbf{$8$ speakers} in the training data set).}
\label{fig:SDR8}
\end{figure*}

Figures \ref{fig:SDR2}, \ref{fig:SDR4}, and \ref{fig:SDR8} show the results of the speech separation performance of the conventional DC models and the proposed X-DC model under the above experimental conditions. From these results, we see that the proposed X-DC could achieve speech separation performance comparable to that of the conventional DC in some settings, while providing an interpretation of the mask estimation mechanism. One possible reason for this is that we have incorporated the smooth spectrogram structure of a speech signal into the output $\tilde{H}$ by adding the penalty term for the difference between $\tilde{H}$ and input spectrogram $X$, as we have described in Section \ref{sec:X_DC}. In the conventional DC models, such ``unnaturalness'' of the output masked spectrogram is not penalized during the training phase, which would have resulted in the lower SDR with some test input signals. On the other hand, in some cases [e.g., $8$ speakers in the training data set, known pair, (m, f)], the proposed X-DC could not achieve as high performance as the conventional DC. In the proposed X-DC, the convolution kernel $W$ was trained with the non-negative constraint, which might have led to the vanishing gradient. Developing a more sophisticated training method to avoid such a problem would be important future work.

\begin{figure*}[!t]
\centering
\includegraphics[width=\hsize]{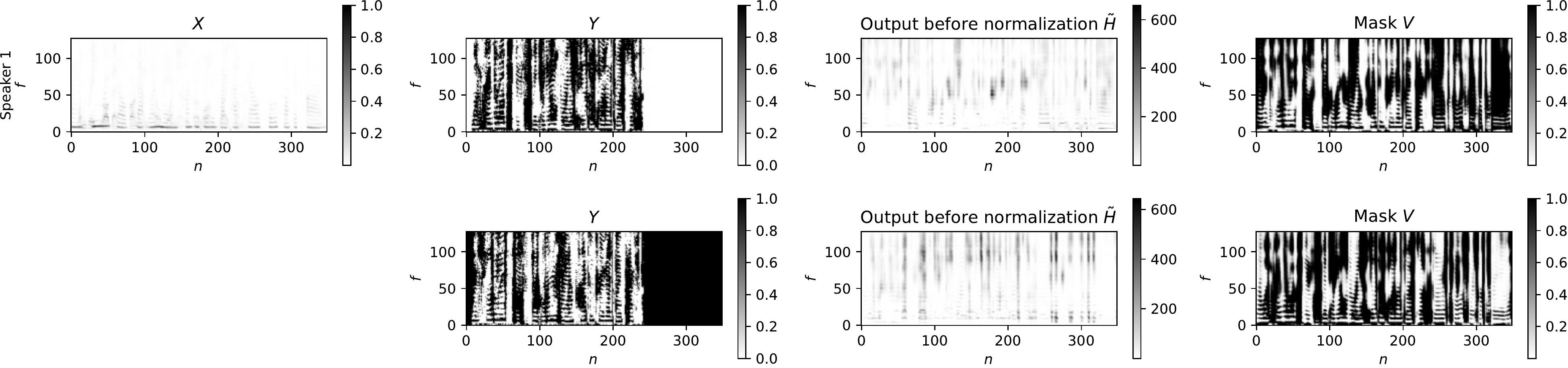}\vspace{-2mm}
\caption{Input spectrogram $X$ to the proposed X-DC model, true speaker label $Y$, output $\tilde{H}$ before normalization of the trained X-DC model, and the Wiener mask $V$, from left to right, respectively (\textbf{$2$ speakers} in the training data set, \textbf{bdl} and \textbf{clb}). The upper and lower figures, respectively, show the results for speakers $1$ and $2$. Note that the results for $Y$ and $V$ are listed in random order (i.e., top or bottom) from the permutation-invariance property of the X-DC.}
\label{fig:mask_bdl_clb}
\end{figure*}

\begin{figure*}[!t]
\centering
\includegraphics[width=\hsize]{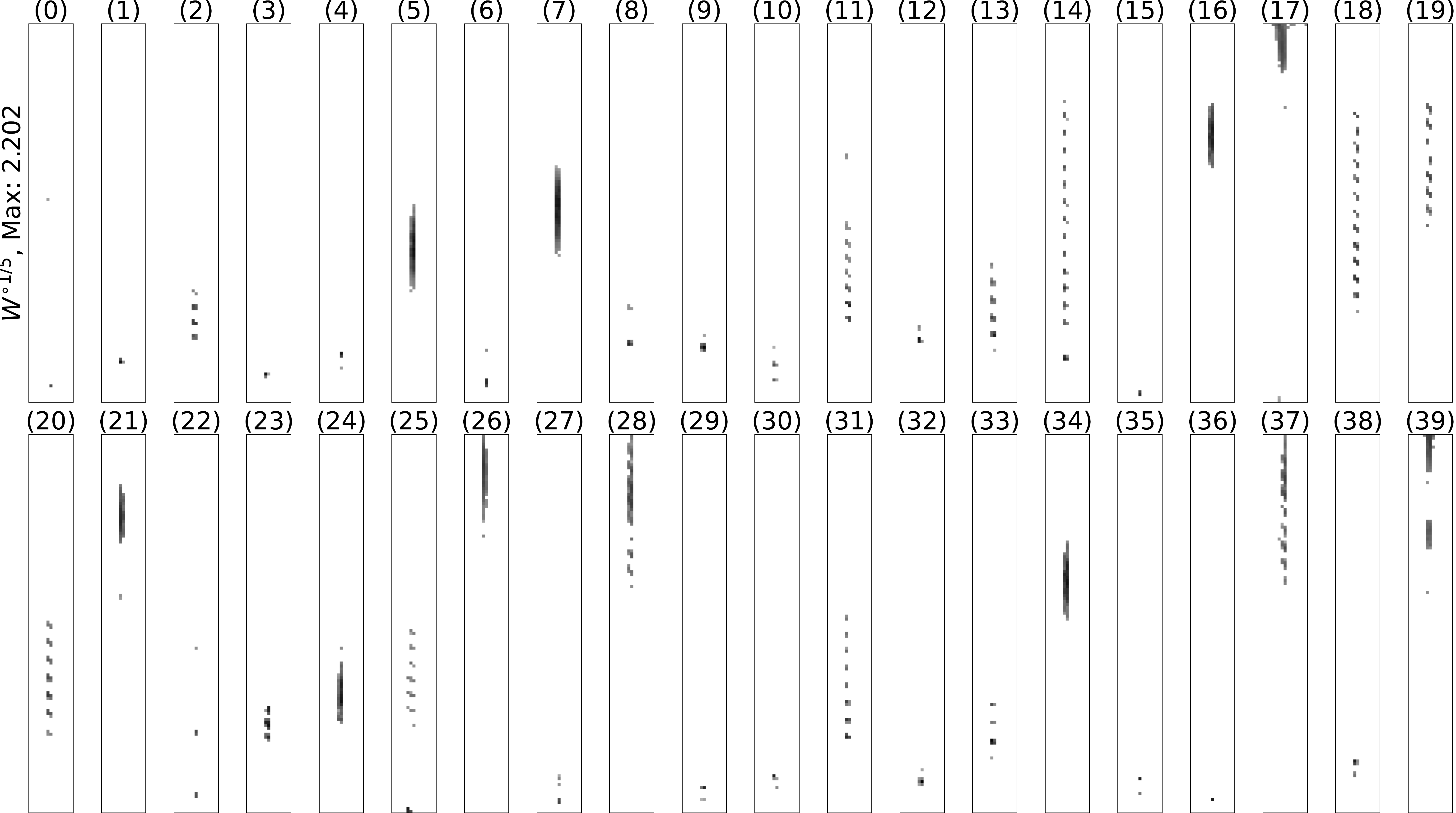}
\caption{$40$ spectrogram templates $W$ of the trained X-DC model (\textbf{$2$ speakers} in the training data set, \textbf{bdl} and \textbf{clb}). Each spectrogram template represents a short-time feature of an input spectrogram, which is used for Wiener mask estimation. Each one has the size of $F \times M$, where the row size $F$ is equal to the number of frequency bins of an input spectrogram, and the column size $M$ is the frame width of a convolutional kernel. For visibility, we plotted $W^{\circ \frac{1}{5}}$, where $\circ$ represents an element-wise power.}\vspace{5mm}
\label{fig:templates}
\end{figure*}

\begin{figure*}[p]
\centering
\includegraphics[width=\hsize]{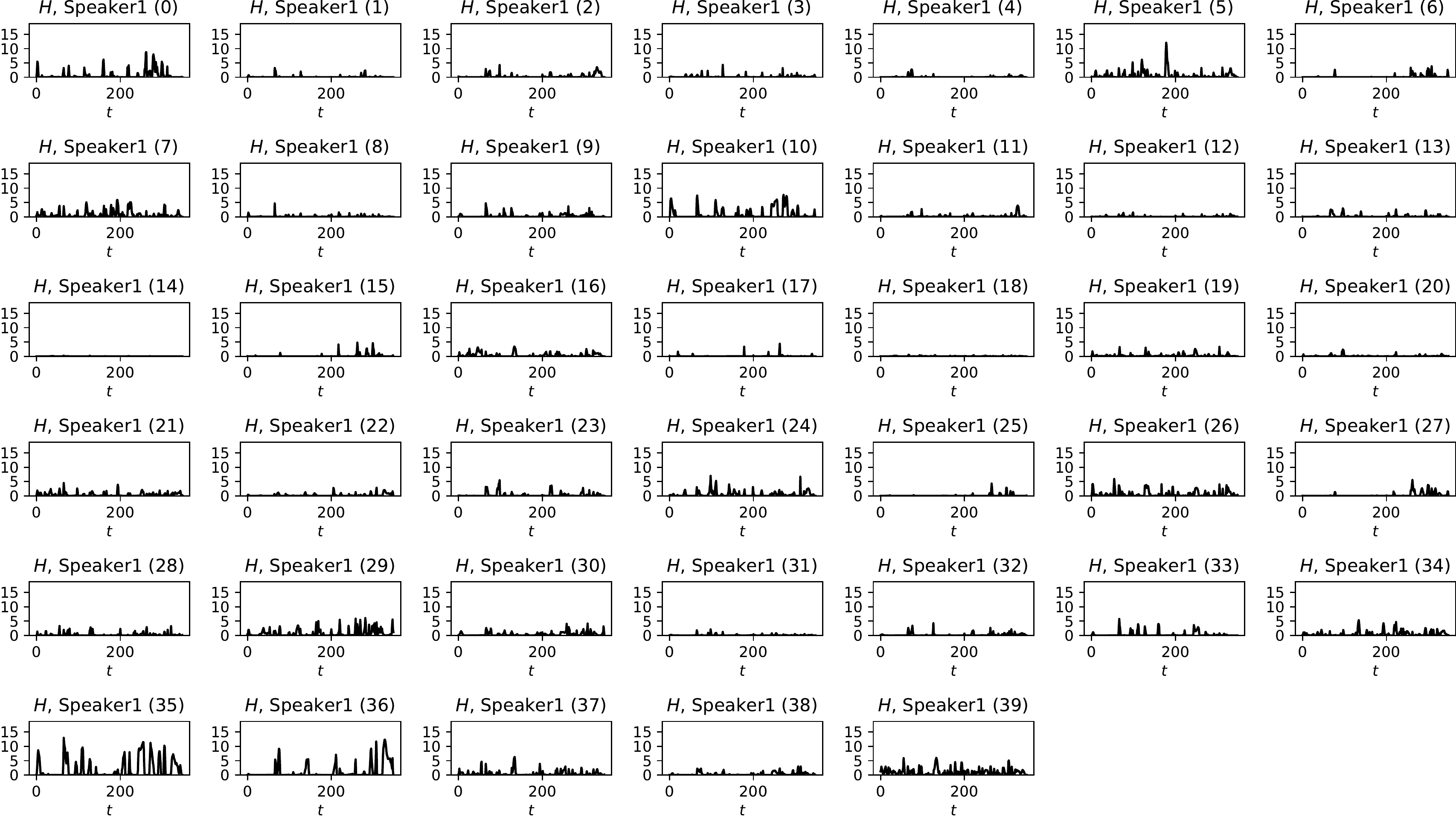}\vspace{-2mm}
\caption{Temporal changes in activation weights $H$ of all the spectrogram templates for \textbf{speaker $1$} (\textbf{$2$ speakers} in the training data set, \textbf{bdl} and \textbf{clb}). The numbers in the parentheses correspond to the indices of the spectrogram templates $W$ in Figure \ref{fig:templates}. In each figure, the horizontal and vertical axes, respectively, show the time bins of an input spectrogram with the size of $N$ and the activation weights of the corresponding spectrogram template.}
\label{fig:activation_1}
\end{figure*}

\begin{figure*}[p]
\centering
\includegraphics[width=\hsize]{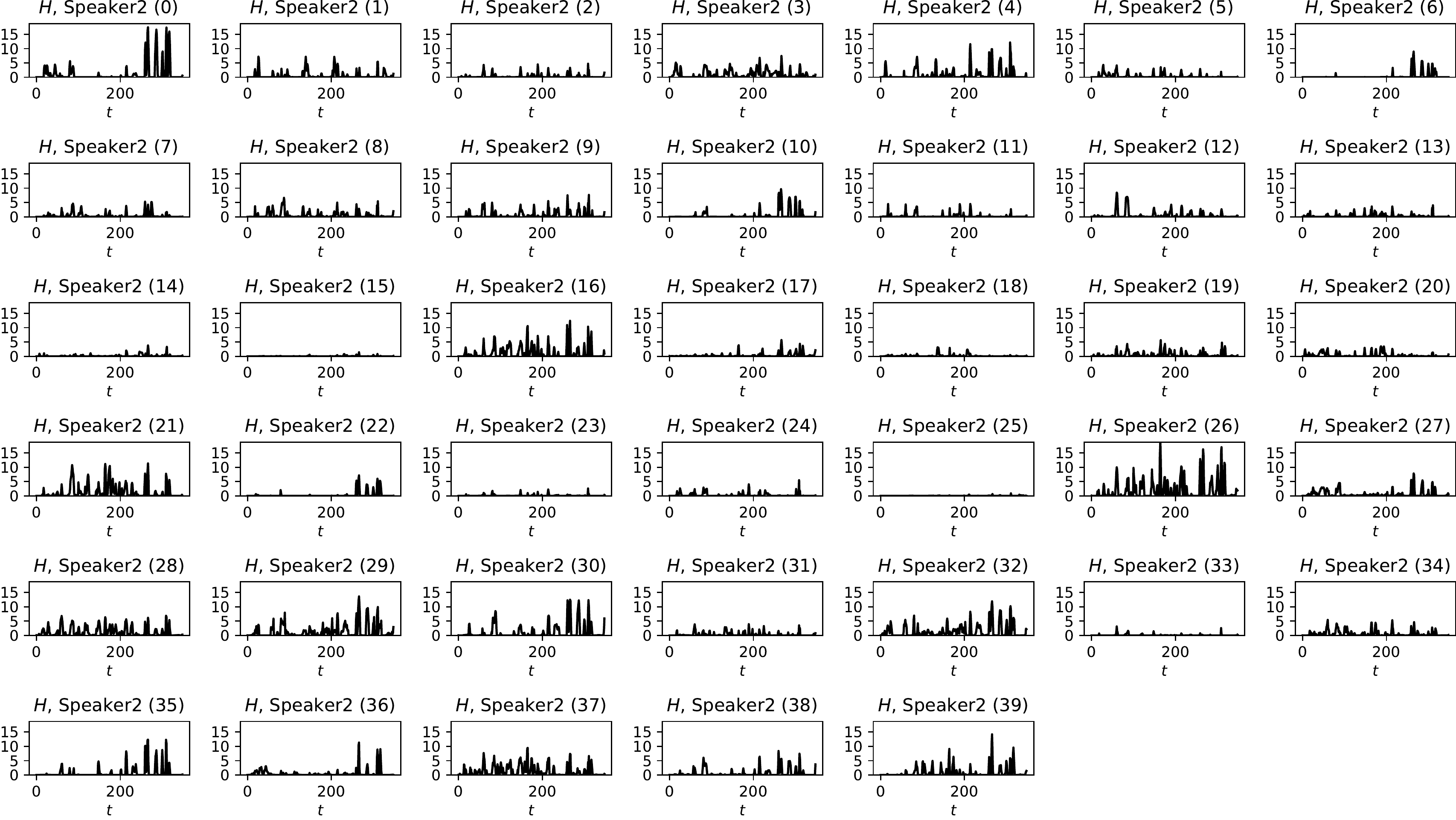}\vspace{-2mm}
\caption{Temporal changes in activation weights $H$ of all the spectrogram templates for \textbf{speaker $2$} (\textbf{$2$ speakers} in the training data set, \textbf{bdl} and \textbf{clb}).}
\label{fig:activation_2}
\end{figure*}

Figure \ref{fig:mask_bdl_clb} shows the input spectrogram $X$ to the proposed X-DC model, true speaker label $Y$, output $\tilde{H}$ before normalization of the trained X-DC model, and the estimated Wiener mask $V$, where there were $2$ speakers (bdl and clb) in the training data set. Figures \ref{fig:templates}, \ref{fig:activation_1}, and \ref{fig:activation_2}, respectively, show the trained spectrogram templates $W$ of the proposed X-DC and the temporal changes in activation weights $H$ of the spectrogram templates in the above setting (Figure \ref{fig:activation_1} and \ref{fig:activation_2}, respectively, correspond to the results of speakers $1$ and $2$). Interestingly, from Figure \ref{fig:templates}, we see that the some of the trained spectrogram templates capture \textbf{harmonic structures} (i.e., features that are distributed over almost equal intervals along a frequency axis) of an input spectrogram for Wiener mask estimation. Note that we have \textit{not} incorporated any prior knowledge about such harmonic structures into the proposed X-DC model, unlike the methods in previous studies \cite{Duan2008, Rabiee2012}. During training, the proposed X-DC model automatically learned to use these templates as informative features. 

%---
%---

%---
%---

%---

We also provide the results of additional experiments in the following appendices: In Appendix \ref{sec:ap_random_weights}, we tried multiplying not a deterministic but random weight to the spectrogram of each speaker before generating the mixed speech signal in the test time. Next, in Appendix \ref{sec:ap_same_D}, we compared the performances of the conventional DC models (i.e., BLSTM-DC \cite{Hershey2016} and conv-DC \cite{Li2018}) by setting their dimensions $D$ of the embedding space at the same value. Third, in Appendix \ref{sec:ap_xdc_M_J}, we checked the changes in performance of the proposed X-DC model with teh different settings of $M$ and $J$, which represent the frame width of a convolutional kernel and the number of spectrogram templates, respectively. In Appendix \ref{sec:ap_16speakers}, we used larger training data set with $16$ different speakers from the CMU Arctic speech databases \cite{arctic2004} and L2-ARCTIC \cite{arctic2018}, and compared the test performance of the proposed and conventional models. In Appendix \ref{sec:ap_potential_speakers}, we checked the performance of the proposed X-DC when setting the maximum potential number of speakers $I$ at larger values than the true one. Finally, in Appendix \ref{sec:ap_danet}, we compared X-DC with DANet \cite{Chen2017}, which is also an end-to-end permutation-free speech separation method that uses a soft mask to perform separation.
%================================================================================

\section{Conclusion}
\label{sec:conclusion}

In this paper, we proposed a new explainable deep clustering (X-DC) model, which extends the original DC model \cite{Li2018} to provide interpretation for speech separation mechanism. To develop such an interpretable model while exploiting the high predictive performance of a neural network, we constructed a network architecture such that its output Wiener mask is computed as a scaled sum of convolutions of short-term spectrogram templates and their activations, both of which are constrained to be non-negative. Experimental results showed that the proposed X-DC model could achieve accuracy comparable to that of the original DC model, and some of the trained spectrogram templates captured the harmonic structures of an input spectrogram, even though we did not incorporate any prior knowledge about such harmonic structures into the proposed X-DC model. 

Recently, many extensions of DC have been proposed. These include the Chimera Network \cite{Luo2017}, which learns to perform mask inference and TF embedding simultaneously, a phase-aware extension \cite{Wang2018a}, and a multi-channel extension \cite{Wang2018b}. We believe that these extended versions can also benefit from the idea proposed in this paper by having model adaptation capabilities. This would be an interesting direction for future work. Another challenge is to consider the reverberant conditions, which may be different between the training and test phases. This problem would be alleviated by incorporating some model adaptation mechanism into the model, which we plan to present in a follow-up paper. 

%================================================================================

\section*{Acknowledgments}

This work was supported by JST CREST Grant Number JPMJCR19A3, Japan. 

%================================================================================

\clearpage
\begin{appendices}
\section{Experiment using the test input mixture generated by random weight}
\label{sec:ap_random_weights}

In the test time in Section \ref{sec:experiment}, we used the deterministic weight of one for the spectrogram of each speaker, when generating the spectrogram of a mixed speech signal. Here, we tried using a random weight in the test phase, instead of such a deterministic one, by generating it from the uniform distribution on $[0, 1)$. Aside from the spectrogram weight, we used the same settings as in Section \ref{sec:experiment}. 

\begin{figure*}[p]
\centering
\includegraphics[width=0.95\hsize]{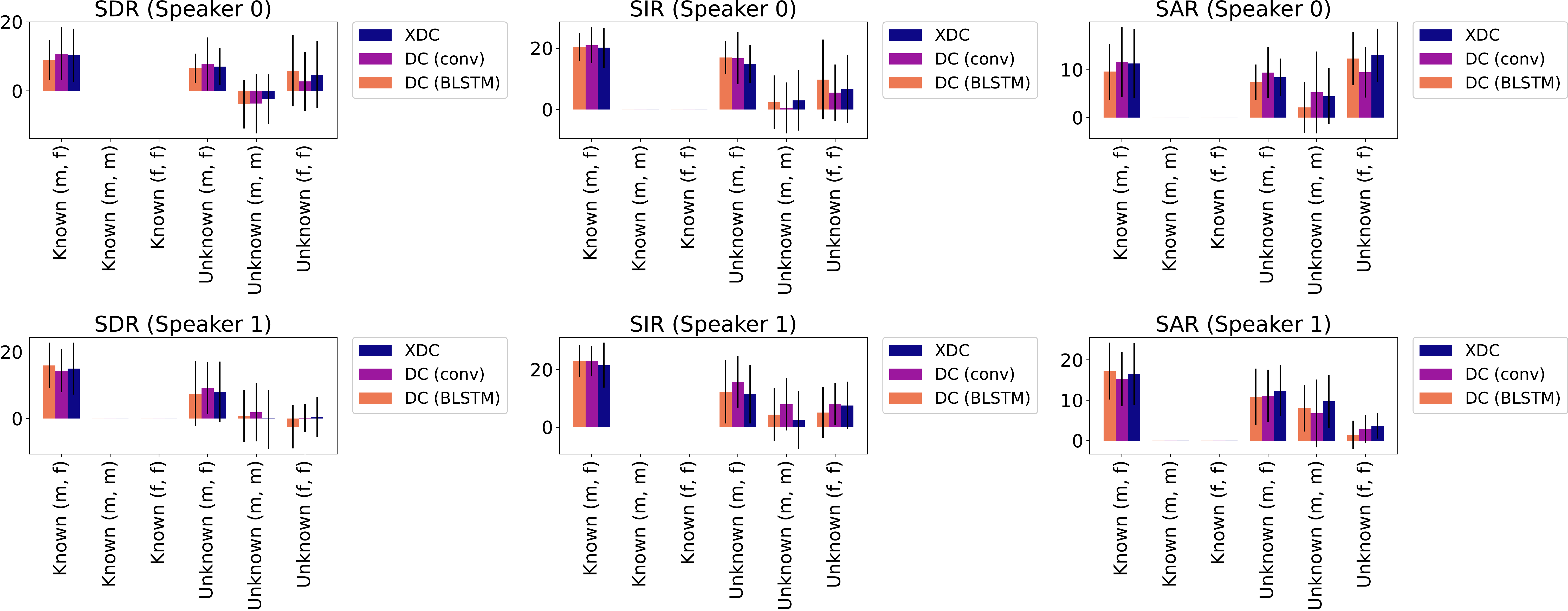}\vspace{-2mm}
\caption{Comparison of speech separation performance (\textbf{$2$ speakers} in the training data set, \textbf{random weight}).}\vspace{5mm}
\label{fig:SDR2_random_weights}
\includegraphics[width=0.95\hsize]{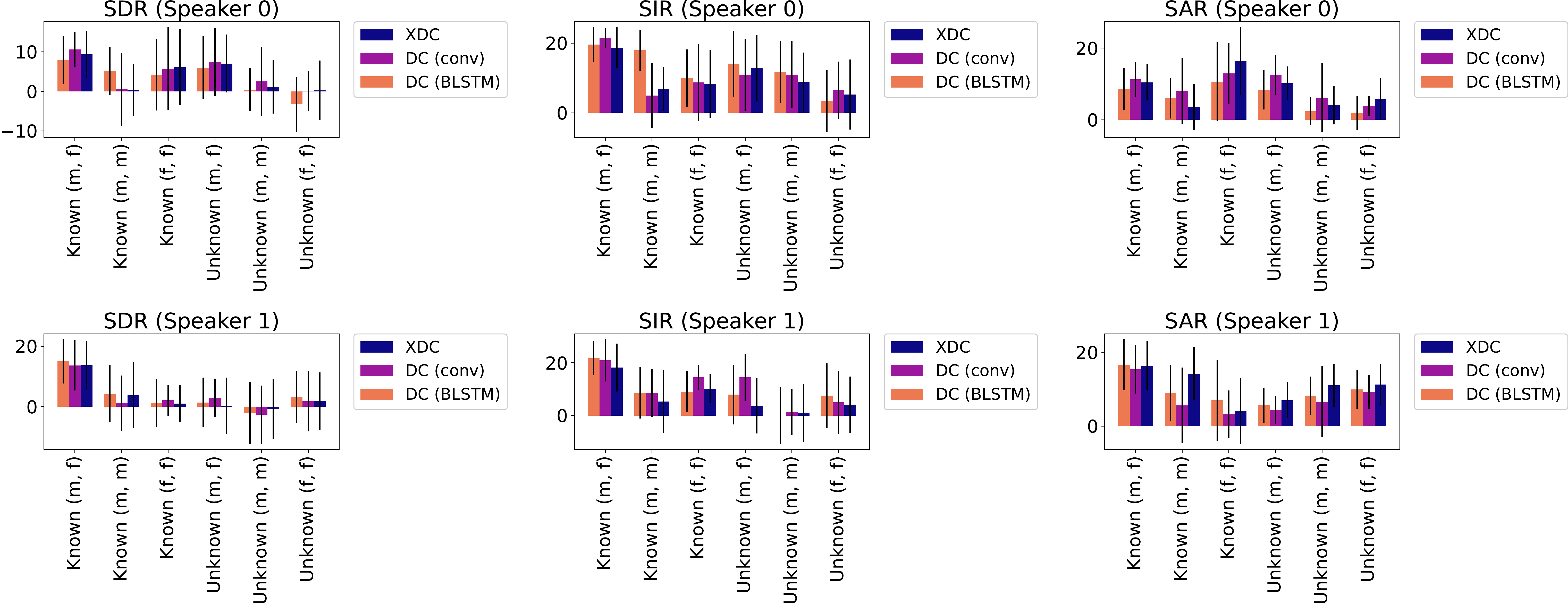}\vspace{-2mm}
\caption{Comparison of speech separation performance (\textbf{$4$ speakers} in the training data set, \textbf{random weight}).}\vspace{5mm}
\label{fig:SDR4_random_weights}
\includegraphics[width=0.95\hsize]{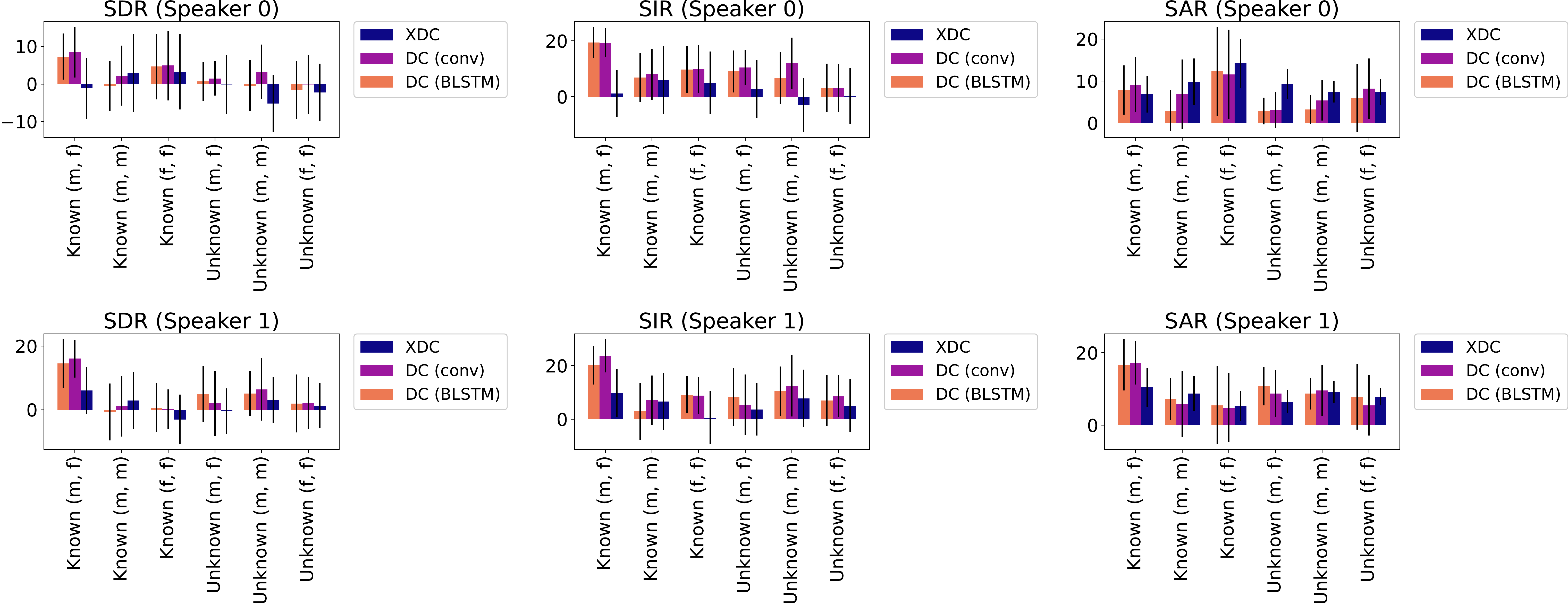}\vspace{-2mm}
\caption{Comparison of speech separation performance (\textbf{$8$ speakers} in the training data set, \textbf{random weight}).}
\label{fig:SDR8_random_weights}
\end{figure*}

Figures \ref{fig:SDR2_random_weights}, \ref{fig:SDR4_random_weights}, and \ref{fig:SDR8_random_weights}, respectively, show the speech separation performances of the three models (i.e., X-DC, conv-DC, and BLSTM-DC) when using the training data set with $2$, $4$, and $8$ different speakers. By comparing these figures with Figures \ref{fig:SDR2}, \ref{fig:SDR4}, and \ref{fig:SDR8}, we see that the variance of the test result increased with a random weight. 

%---
%---

%================================================================================

\section{Experiment for comparing BLSTM-DC and conv-DC with the same $D$ setting}
\label{sec:ap_same_D}

We also checked the performances of the conventional DC models (i.e., BLSTM-DC and conv-DC) by setting their dimensions $D$ of the embedding space at the same value. Specifically, we tried the following two patterns: $D = 20$ and $D = 90$, which we used for the BLSTM-DC and conv-DC, respectively, in Section \ref{sec:experiment}. Aside from $D$, we used the same settings (including all the hyperparameters for the proposed X-DC) as in Section \ref{sec:experiment}. 

\begin{figure*}[p]
\centering
\includegraphics[width=0.95\hsize]{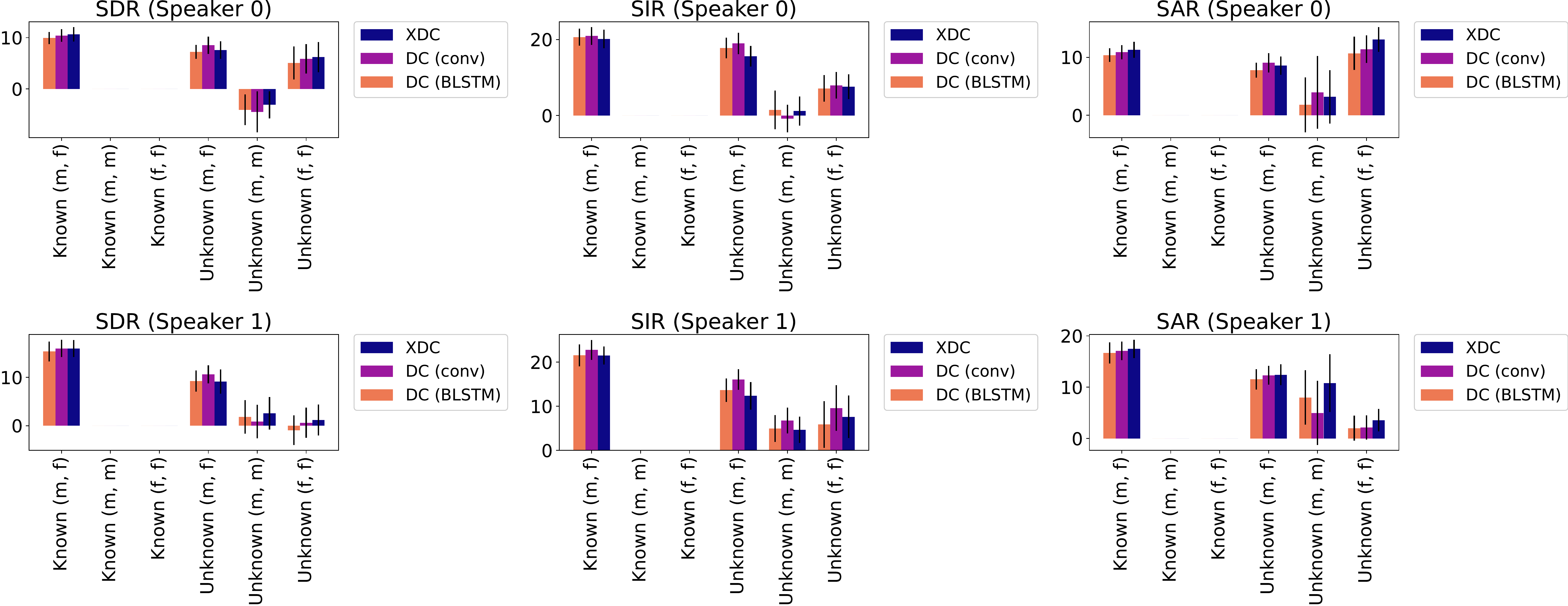}\vspace{-2mm}
\caption{Comparison of speech separation performance (\textbf{$2$ speakers} in the training data set, \textbf{$D = 20$}).}\vspace{5mm}
\label{fig:SDR2_D20}
\includegraphics[width=0.95\hsize]{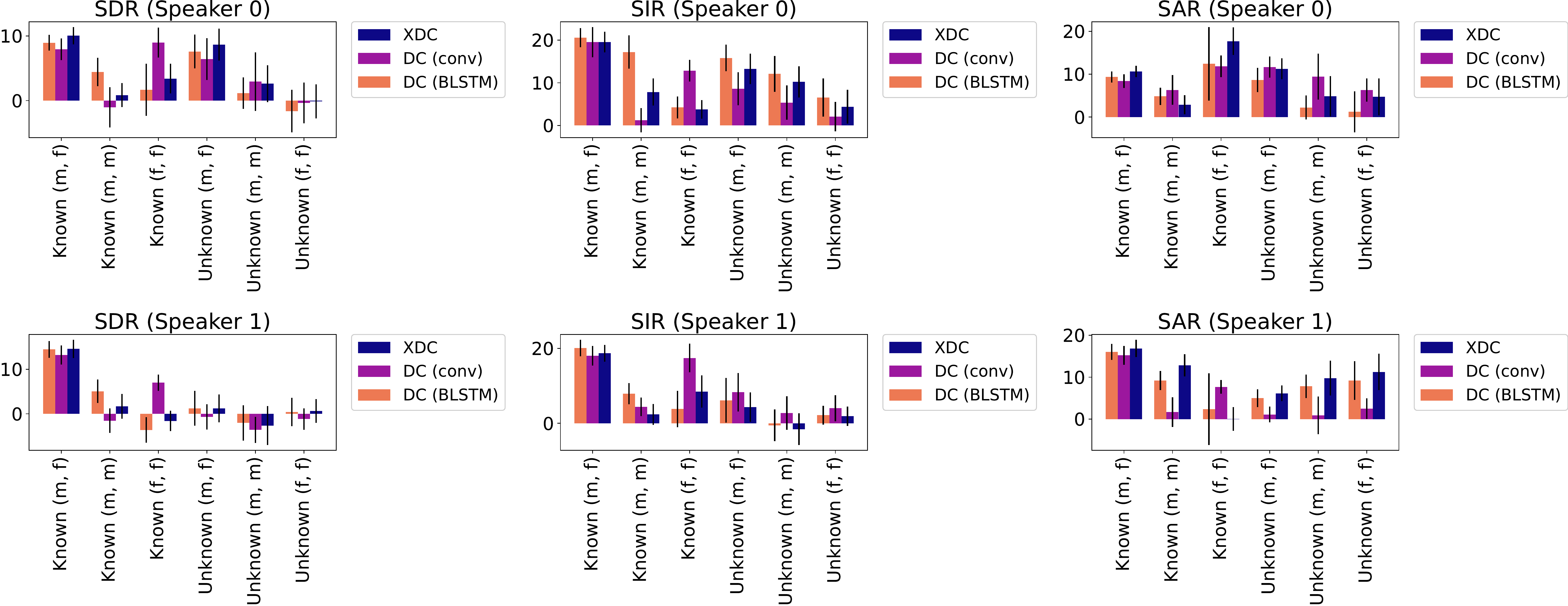}\vspace{-2mm}
\caption{Comparison of speech separation performance (\textbf{$4$ speakers} in the training data set, \textbf{$D = 20$}).}\vspace{5mm}
\label{fig:SDR4_D20}
\includegraphics[width=0.95\hsize]{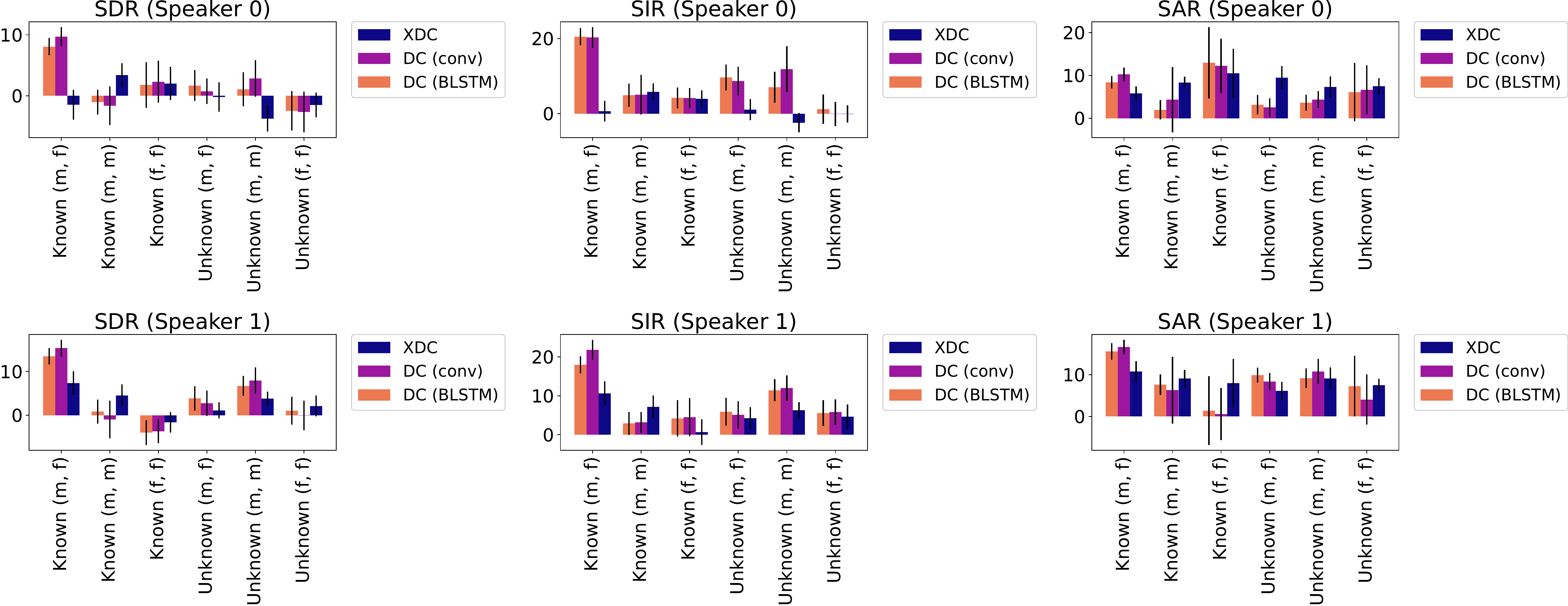}\vspace{-2mm}
\caption{Comparison of speech separation performance (\textbf{$8$ speakers} in the training data set, \textbf{$D = 20$}).}
\label{fig:SDR8_D20}
\end{figure*}

\begin{figure*}[p]
\centering
\includegraphics[width=0.95\hsize]{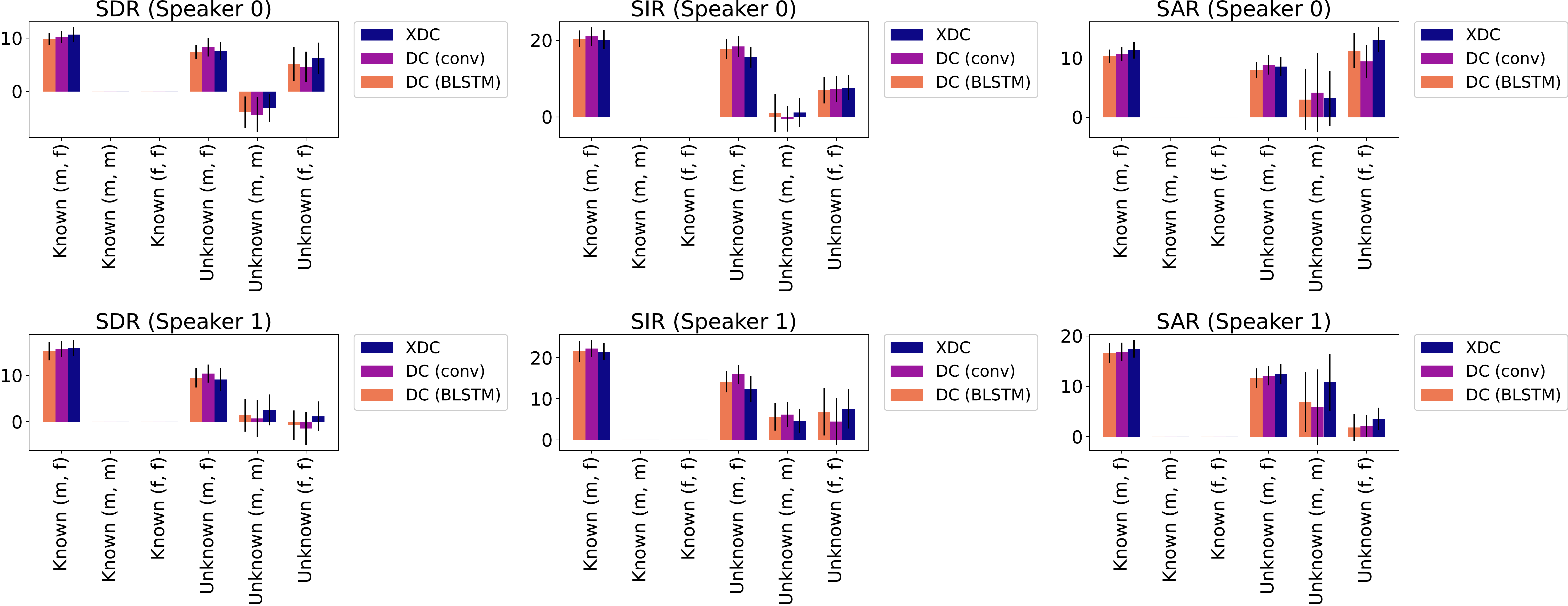}\vspace{-2mm}
\caption{Comparison of speech separation performance (\textbf{$2$ speakers} in the training data set, \textbf{$D = 90$}).}\vspace{5mm}
\label{fig:SDR2_D90}
\includegraphics[width=0.95\hsize]{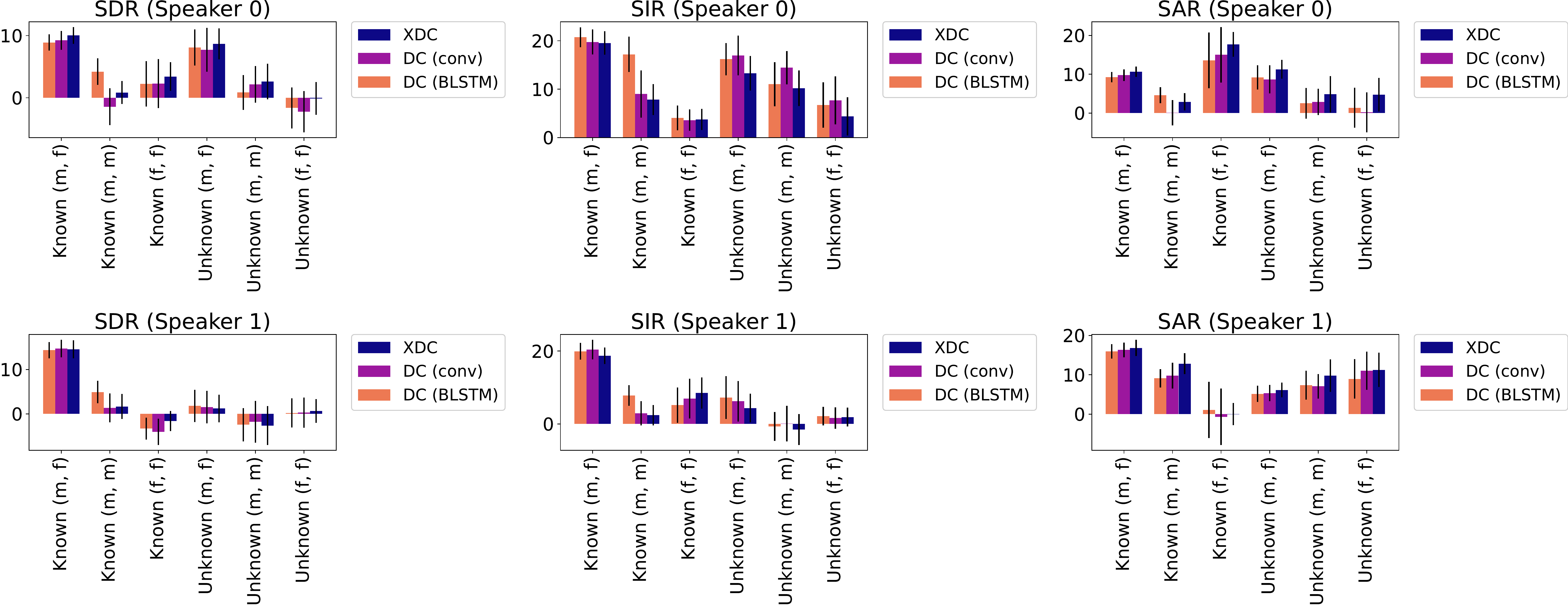}\vspace{-2mm}
\caption{Comparison of speech separation performance (\textbf{$4$ speakers} in the training data set, \textbf{$D = 90$}).}\vspace{5mm}
\label{fig:SDR4_D90}
\includegraphics[width=0.95\hsize]{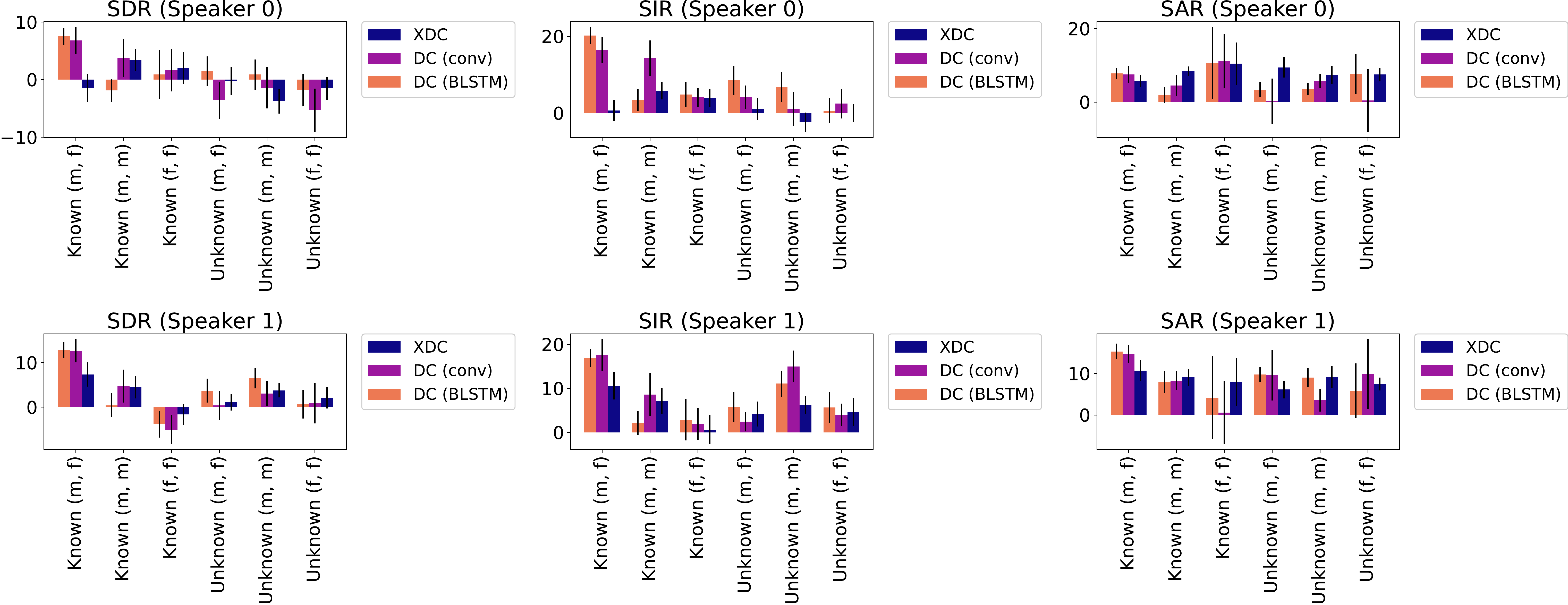}\vspace{-2mm}
\caption{Comparison of speech separation performance (\textbf{$8$ speakers} in the training data set, \textbf{$D = 90$}).}
\label{fig:SDR8_D90}
\end{figure*}

Figures \ref{fig:SDR2_D20}, \ref{fig:SDR4_D20}, and \ref{fig:SDR8_D20}, show the results for the setting of $D = 20$, and Figures \ref{fig:SDR2_D90}, \ref{fig:SDR4_D90}, and \ref{fig:SDR8_D90}, show those for $D = 90$. When the training data set contains exactly $2$ different speakers (i.e., Figures \ref{fig:SDR2_D20} and \ref{fig:SDR2_D90}), the setting of $D$ did not have a large impact on the relative speech separation performances of the two models. On the other hand, when there were $4$ or $8$ different speakers in the training data set, the relative performances of these models changed more drastically according to the setting of $D$. From these figures, we see that the BLSTM-DC was more robust with the change of $D$ than the conv-DC. 

%---
%---

%---
%---

%================================================================================

\section{Experiment with different $M$ and $J$ settings in the X-DC}
\label{sec:ap_xdc_M_J}

To check the effect of changing the frame width of a convolutional kernel $M$ and the number of spectrogram templates $J$ of the porposed X-DC model on the speech separation performance, we tried multiple combinations of their settings. Aside from the values of $M$ and $J$, we used the same settings as in Section \ref{sec:experiment}. 

\begin{figure*}[!t]
\centering
\includegraphics[width=\hsize]{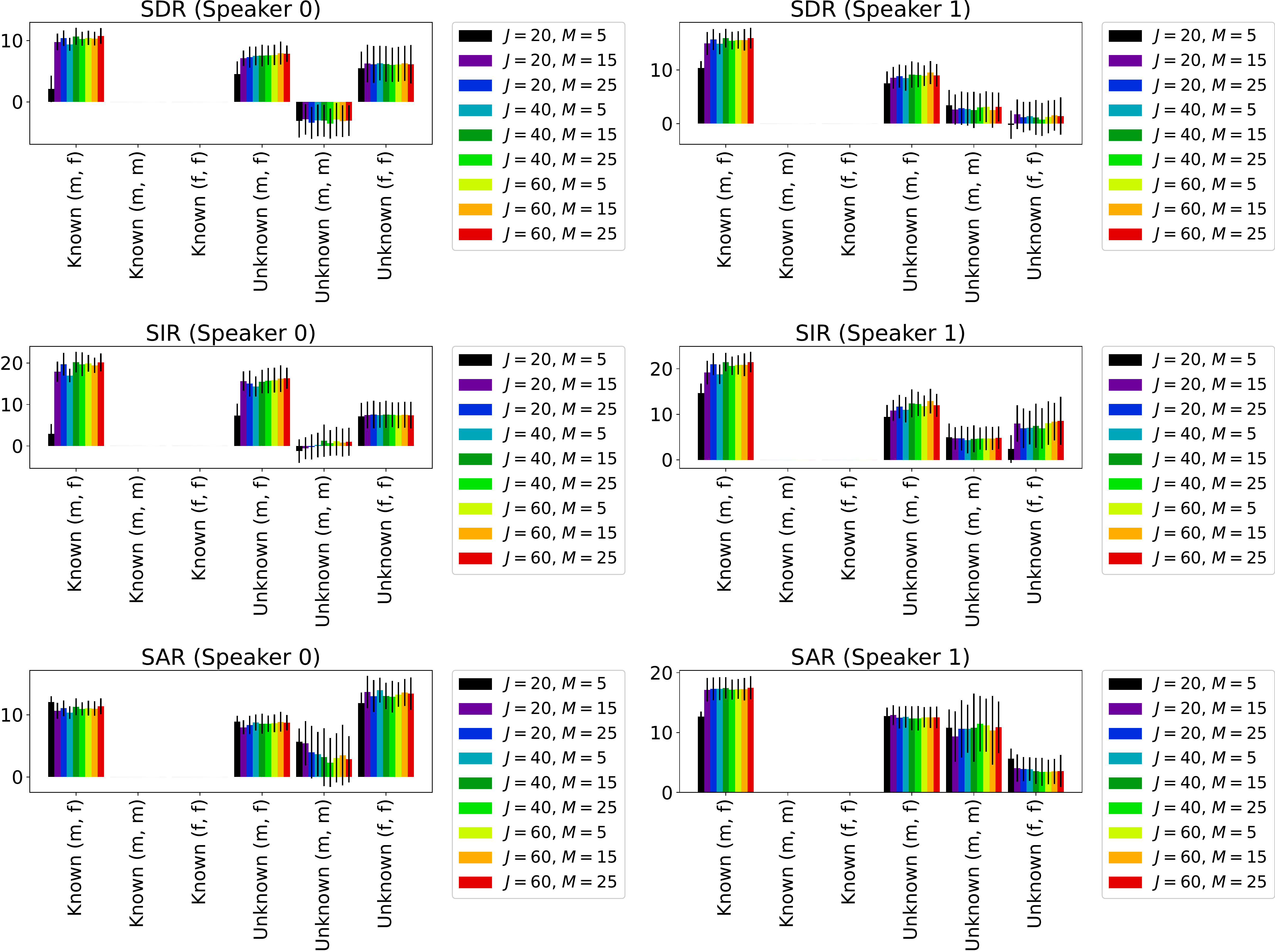}\vspace{-2mm}
\caption{Comparison of speech separation performance between different settings of $J$ and $M$ (\textbf{$2$ speakers} in the training data set).}\vspace{5mm}
\label{fig:SDR2_JM}
\end{figure*}

\begin{figure*}[!t]
\centering
\includegraphics[width=\hsize]{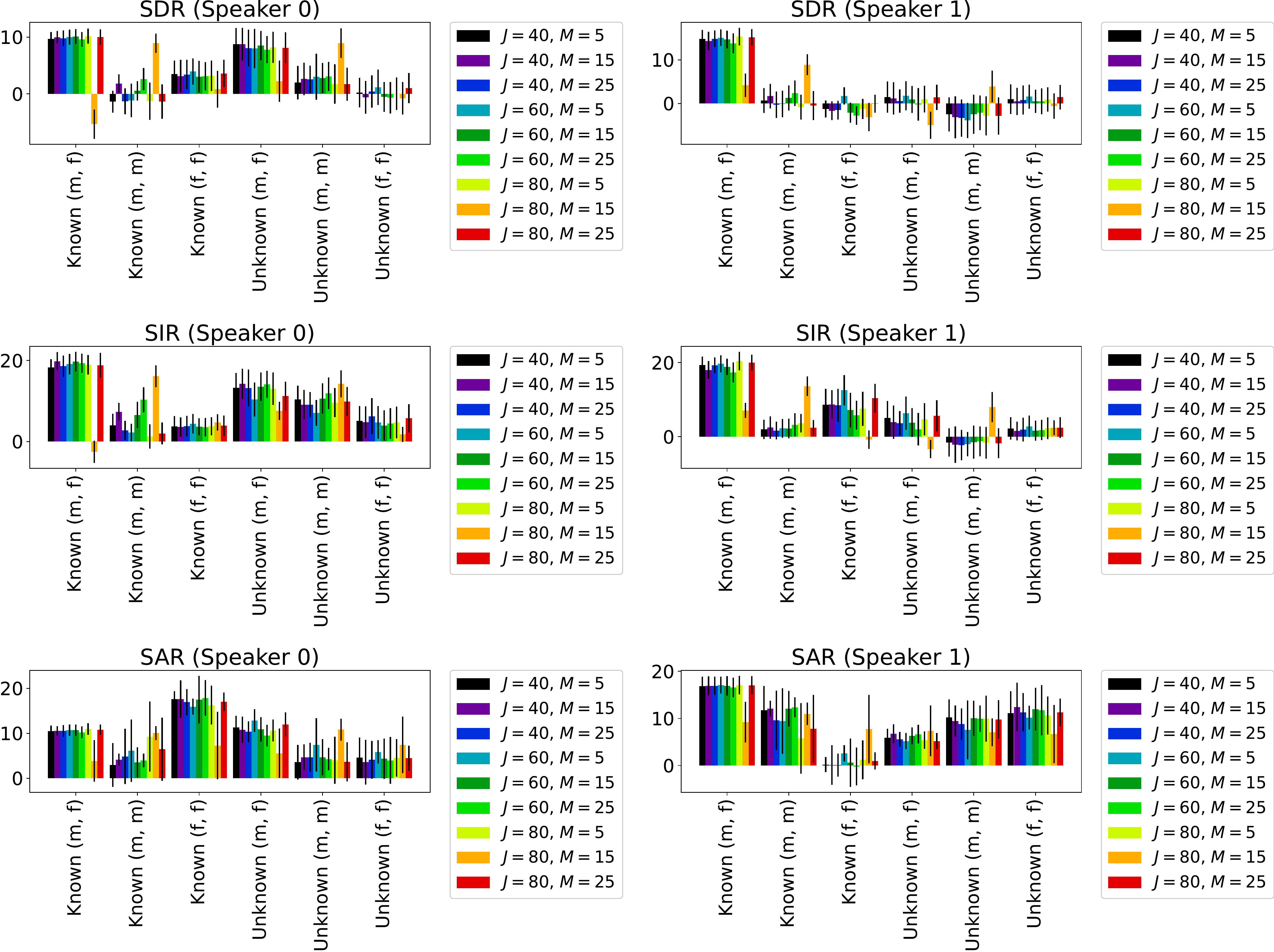}\vspace{-2mm}
\caption{Comparison of speech separation performance between different settings of $J$ and $M$ (\textbf{$4$ speakers} in the training data set).}\vspace{5mm}
\label{fig:SDR4_JM}
\end{figure*}

\begin{figure*}[!t]
\centering\includegraphics[width=\hsize]{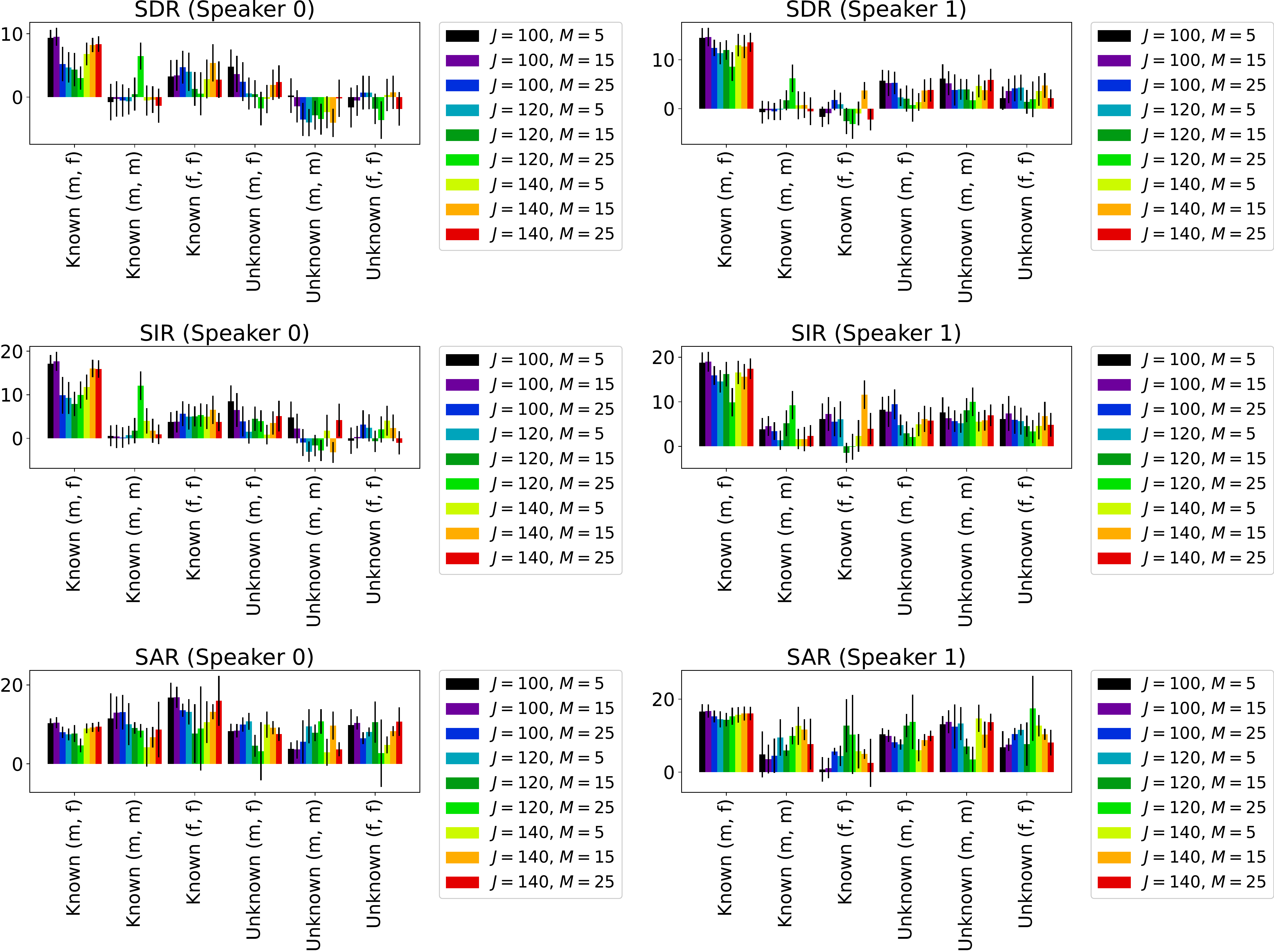}\vspace{-2mm}
\caption{Comparison of speech separation performance between different settings of $J$ and $M$ (\textbf{$8$ speakers} in the training data set).}
\label{fig:SDR8_JM}
\end{figure*}

Figures \ref{fig:SDR2_JM}, \ref{fig:SDR4_JM}, and \ref{fig:SDR8_JM} show the speech separation performances of the proposed X-DC model under nine settings of $(J, M)$ (the combinations of three settings of $J$ and three settings of $M$). From Figure \ref{fig:SDR2_JM}, when the training data set contains $2$ types of speakers, the SDR, SIR, and SAR did not change significantly with the setting of $M$ and $J$ in most cases. In case where the training data set contains $4$ or $8$ different speakers, from Figures \ref{fig:SDR4_JM} and \ref{fig:SDR8_JM}, we see that the performance of the X-DC was more strongly affected by the setting of $(M, J)$. For instance, in Figure \ref{fig:SDR4_JM}, the setting of $(M, J)=(15, 80)$ yielded lower performance with a set of known (male, female) speakers than the other settings of $(M, J)$, while it achieved higher performance with a set of known (male, male) speakers than the other settings of $(M, J)$ in most cases.

%---
%---

%================================================================================

\section{Experiment using larger training data set with $16$ different speakers}
\label{sec:ap_16speakers}

Here, we compare the performances of the proposed and conventional models by using a larger training data set than in Section \ref{sec:experiment}. Specifically, we tried the case where the training data set contained $16$ different speakers of the CMU Arctic speech databases \cite{arctic2004} and L2-ARCTIC \cite{arctic2018}. 

%---
\begin{table*}[t]
\centering
\caption{Experimental settings of the speakers in the training, validation, and test data sets. ``Known'' shows that the pair of speakers in the test data set has also been in the training data set, and ``unknown'' shows otherwise. M and F, respectively, stand for male and female speakers.}\vspace{0.02\hsize}
\begin{tabular}{|c||p{0.25\hsize}|c|c|c|c|} \hline 
\rowcolor[rgb]{0.93, 0.93, 0.93} \multicolumn{1}{|c||}{\textbf{Setting \#}} & \multicolumn{1}{c|}{\textbf{Training data}} & \multicolumn{1}{c|}{\textbf{Validation data}} & \multicolumn{3}{c|}{\textbf{Test data}} \\ \hline \hline
$19$ & \multirow{6}{*}{\parbox{\hsize}{\textbf{$16$ speakers} (bdl, clb, rms, slt, aba, zhaa, bwc, lxc, asi, svbi, hkk, hjk, ybaa, mbmps, rrbi, ncc)}} & \multirow{6}{*}{bdl, clb} & \multirow{3}{*}{Known} & M/F & bdl, clb \\
$20$ & & & & M/M & bdl, rms \\
$21$ & & & & F/F & clb, slt \\ \cline{4-6}
$22$ & & & \multirow{3}{*}{Unknown} & M/F & ykwk, tni \\
$23$ & & & & M/M & ykwk, erms \\
$24$ & & & & F/F & tni, ydck \\ \hline 
\end{tabular}
\label{tb:setting_speaker16}
\end{table*} 
%---

Table \ref{tb:setting_speaker16} shows the experimental settings of the speakers in the training, validation, and test data sets. We generated these data sets based on the same procedure as in Section \ref{sec:experiment}. For both the validation and test data sets, in any case (i.e., known or unknown speakers and gender combination of the two speakers in the test phase), we used $66$ sets of mixed speech signals of the two speakers shown in Table \ref{tb:setting_speaker16}. 

The detailed numerical settings are as follows: 
\begin{itemize}
\item For the conventional BLSTM-DC model \cite{Hershey2016}, we set the dimension of the embedding space at $D = 20$, the number of hidden cells in each BLSTM layers at $600$, the number of the BLSTM layers at $3$, the learning rate at $10^{-5}$, and the number of epochs for training at $T = 1100$. Under these settings, the number of learnable parameters in the conventional DC model of a BLSTM network is $23,880,160$.
\item For the conventional conv-DC model \cite{Li2018}, we set the dimension of the embedding space at $D = 40$, the number of output channels in the middle convolutional layers at $C = 11$, the number of the middle convolutional layers at $L^{\mathrm{conv}} = 3$ (layers) $\times N^{\mathrm{block}}$ (blocks) with $N^{\mathrm{block}} = 45$, the learning rate at $\eta = 5 \times 10^{-4}$, and the number of epochs for training at $T = 600$. Under these settings, the number of learnable parameters in the conventional DC model of a gated convolutional network is $1,256,658$.
\item For the proposed X-DC model, we set the maximum potential number of speakers at $I = 2$, the regularization hyperparameter at $\lambda = 5 \times 10^{-4}$, the frame width of a convolutional kernel at $M = 15$, the number of spectrogram templates at $J = 50$, the number of output channels in the middle convolutional layers at $C = 188$, the number of layers of ``NMFD part'' at $L^{\mathrm{NMFD}} = 7$, the learning rate at $\eta = 5 \times 10^{-3}$, and the number of epochs for training at $T = 900$. Under these settings, the number of learnable parameters in the proposed X-DC model is $1,197,604$.
\item We used the same common settings as in Section \ref{sec:experiment} for the batch size, sampling rate, window size of the STFT, window shift, number of time bins of an input spectrogram, and training algorithm. The number of utterances in the training data set of each of the eight speakers (that have not been used in Section \ref{sec:experiment}) is as follows: $1,000$ (mbmps, erms), $999$ (ncc, ykwk, tni, ydck), and $998$ (ybaa, rrbi).
\end{itemize}
The above hyperparameter settings (i.e., $T$ for the BLSTM-DC, $D$, $C$, $N^{\mathrm{block}}$, $\eta$, and $T$ for the conv-DC, and $J$, $C$, $L^{\mathrm{NMFD}}$, $\eta$, $T$, and $\lambda$ for the proposed X-DC) were chosen by the hold-out validation with $66$ validation data sets.

\begin{figure*}[!t]
\centering
\includegraphics[width=0.95\hsize]{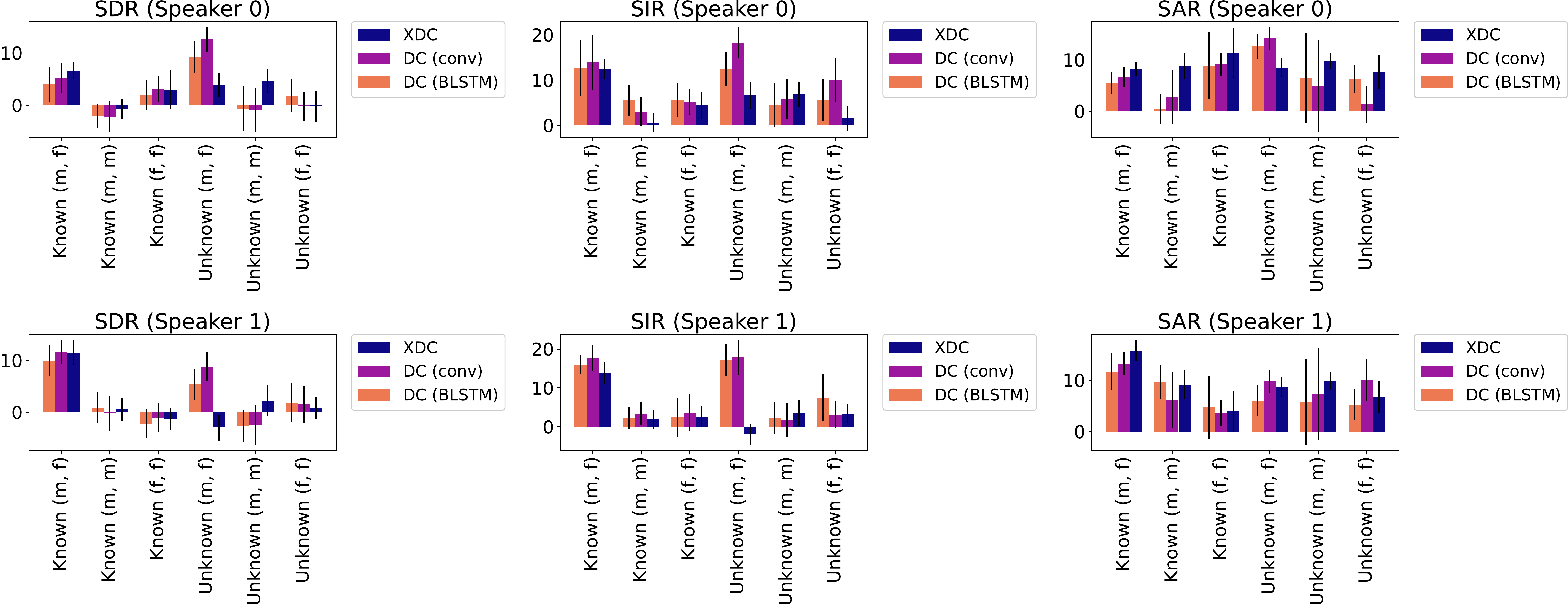}\vspace{-2mm}
\caption{Comparison of speech separation performance (\textbf{$16$ speakers} in the training data set.}
\label{fig:SDR16}
\end{figure*}

Figure \ref{fig:SDR16} shows the results of  the speech separation performance of the proposed and conventional models, where the training data set contains $16$ different speakers. From this figure, we see that in the setting of ``unknown (m, f)'' (i.e., the test input contains male and female speakers who were not included in the training data set), the conv-DC and BLSTM-DC achieved as high performance as in the case of ``known (m, f)'' (i.e., the test input contains male and female speakers who were included in the training data set), except for some cases (e.g., SDR and SAR for Speaker $1$). On the other hand, when the gender combination of the test data set is (male, male) or (female, female), the proposed X-DC maintained as high performance with unknown speakers as with known ones in most settings. 

%---
%---

%================================================================================

\section{Experiment under the condition where the assumed number of masks exceeds the actual number of speakers}
\label{sec:ap_potential_speakers}

The proposed X-DC model enables us to separate an input mixed speech signal into an arbitrary number $I$. Here, $I$ represents the maximum potential number of speakers, and its value is not required to be exactly the same as the true number of speakers $I^{\mathrm{true}}$ in the mixed speech signal. Here, we consider the case of $I^{\mathrm{true}} = 2$ as in Section \ref{sec:experiment}, and check the performance of the proposed X-DC under the multiple settings of $I$, where $I \geq I^{\mathrm{true}}$. In the test case of $I > I^{\mathrm{true}}$, we first obtain $I$ separated spectrograms $\tilde{H}^{(1)}, \dots, \tilde{H}^{(I)}$ using the trained X-DC model, and select $I^{\mathrm{true}}$ spectrograms with the maximum Frobenius norms. Then, for selected $I^{\mathrm{true}}$ spectrograms, we compute the SDR, SIR, and SAR and plot the results. 

\begin{figure*}[!t]
\centering
\includegraphics[width=\hsize]{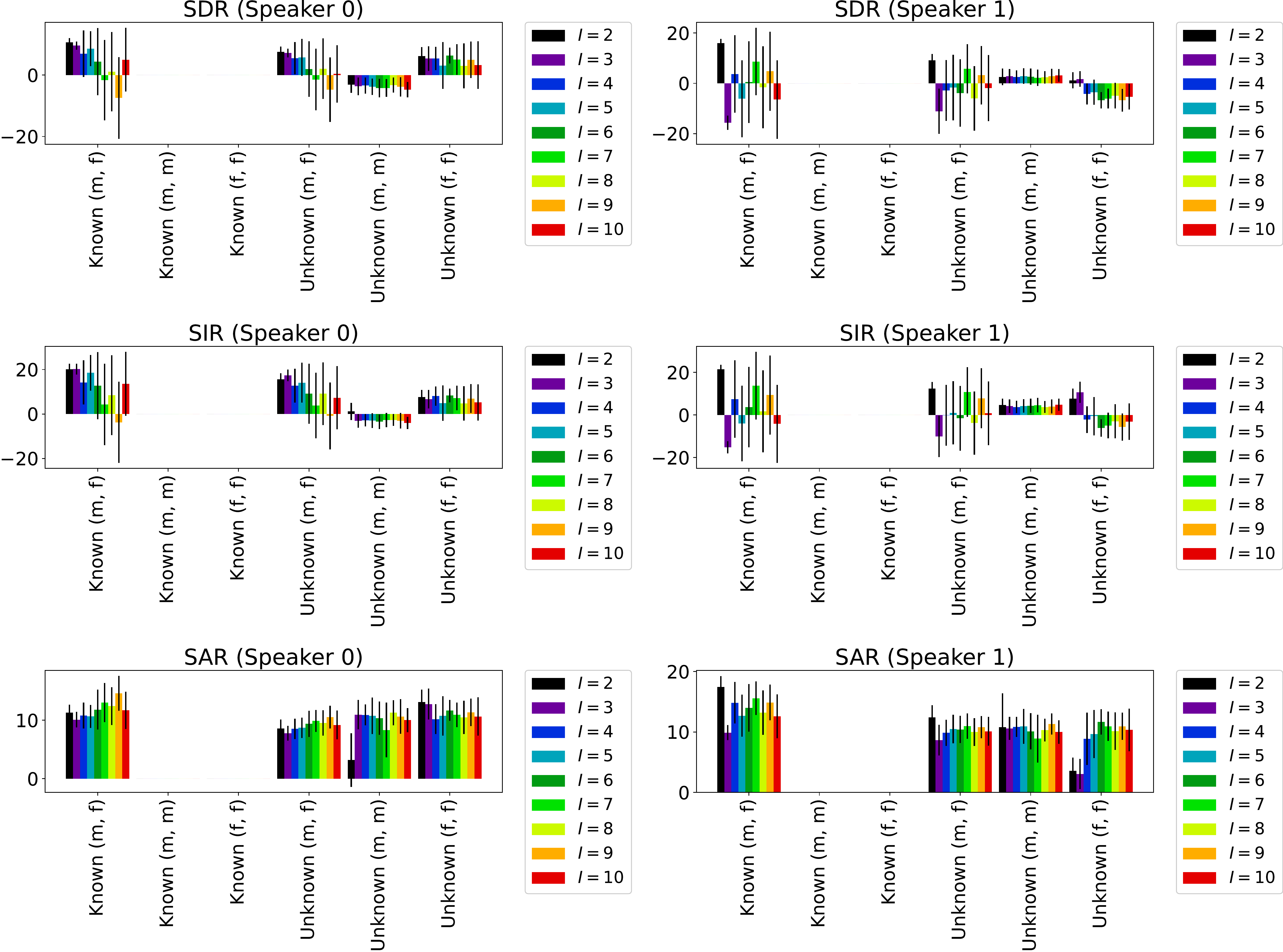}\vspace{-2mm}
\caption{Comparison of speech separation performance between different settings of the maximum potential number of speakers $I$ (\textbf{$2$ speakers} in the training data set).}\vspace{5mm}
\label{fig:SDR_potential_2}
\end{figure*}

\begin{figure*}[!t]
\centering
\includegraphics[width=\hsize]{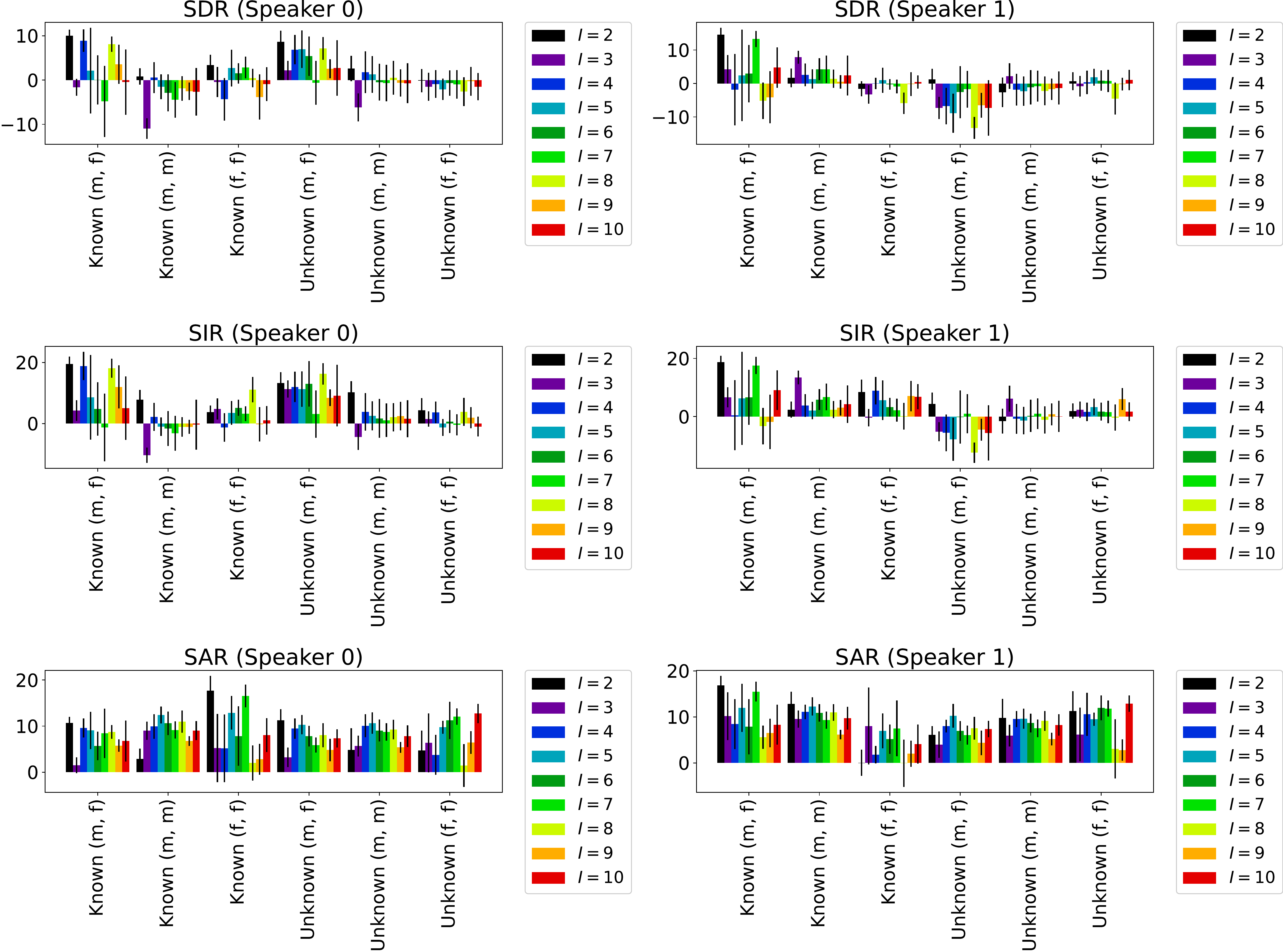}\vspace{-2mm}
\caption{Comparison of speech separation performance between different settings of $I$ (\textbf{$4$ speakers} in the training data set).}\vspace{5mm}
\label{fig:SDR_potential_4}
\end{figure*}

\begin{figure*}[!t]
\centering
\includegraphics[width=\hsize]{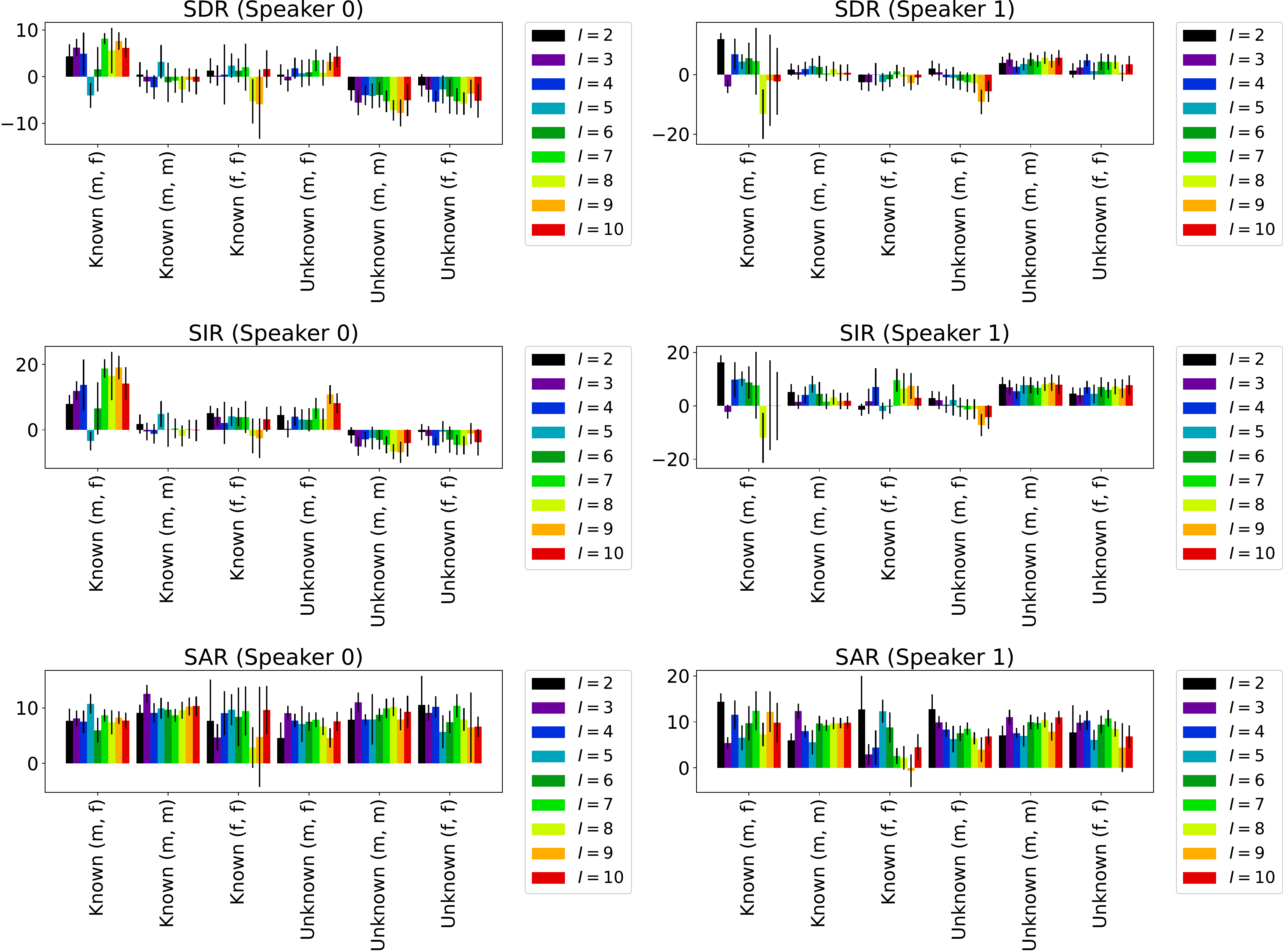}\vspace{-2mm}
\caption{Comparison of speech separation performance between different settings of $I$ (\textbf{$8$ speakers} in the training data set).}
\label{fig:SDR_potential_8}
\end{figure*}

Figure \ref{fig:SDR_potential_2}, \ref{fig:SDR_potential_4}, and \ref{fig:SDR_potential_8}, show the speech separation performances of the proposed X-DC model under the different settings of $I$. From this figure, we see that when the training data set contains $2$ or $4$ different speakers, the models of $I = 2$ or $I = 3$ achieved the best performance in most cases in terms of the SDR. However, in other cases, the setting of $I=2$ (i.e., the maximum potential number of speakers is set at the true number of speakers) did not always yield the best test performance. One possible reason for this result would be that when the input signal could not successfully separated, the ``interference factor'' (i.e., the signal of different speakers from the one being focused on) got smaller by setting redundant potential speakers (i.e., $I \geq 3$). Moreover, in practice, it is often the case that the true number of speakers in an input signal is unknown advance. The proposed X-DC is applicable even in such cases, by setting $I$ at some sufficiently large value. 

%---
%---

%================================================================================

\section{Comparison with DANet}
\label{sec:ap_danet}

While the original DC uses a binary mask to perform separation, the proposed X-DC uses a soft mask. This difference may have had some effect on the SDR scores shown earlier. To evaluate the performance excluding the effect of the choice of mask type, we also compared X-DC with DANet \cite{Chen2017}, an end-to-end permutation-free speech separation method that uses a soft mask to perform separation like X-DC.

Given a magnitude spectrogram $X \in \mathbb{R}^{F \times N}$ of a mixture signal of $I$ sources, DANet learns to obtain a Wiener mask $\hat{m}_{f, n}^{(i)}$ that minimizes the following loss function: 
\begin{align}
&\mathcal{J}^{\mathrm{DANet}} (\theta) = \frac{1}{FNI} \sum_{f = 1}^F \sum_{n = 1}^N \sum_{i = 1}^I \left| X_{f, n} \left( m_{f, n}^{(i)} - \hat{m}_{f, n}^{(i)} \right) \right|, 
\end{align}
where $m_{f, n}^{(i)} \equiv \left( \tilde{X}^{(i)}_{f, n} \right)^2 / \left( X_{f, n}^2 + 10^{-16} \right)$ and $\tilde{X}^{(i)}_{f, n}$ is the magnitude spectrogram of the $i$th speaker. In DANet, the estimated Wiener mask is given by
\begin{align}
\hat{m}_{f, n}^{(i)} = \sigma \left( \sum_{d = 1}^D A_{i, d} v_{F(n-1)+f, d} \right), 
\end{align}
where $A = (A_{i, d})_{1 \leq i \leq I, 1 \leq d \leq D}$ and $V = (v_{k, d})_{1 \leq k \leq FN, 1 \leq d \leq D}$, respectively, are the \textit{attractors} and embeddings of the TF bins, and $\sigma (x) \equiv 1/(1+\exp(-x))$. The embeddings $V$ is obtained by feeding the magnitude spectrogram $X$ of a mixture signal into BLSTM layers: $V = \mathrm{BLSTM} (X)$. As for the attractors $A$, they are given by
\begin{align}
A_{i, d} = \frac{\sum_{k = 1}^{FN} v_{k, d} y_{k, i}}{\sum_{k = 1}^{FN} y_{k, i} + 10^{-8}}, 
\end{align}
where $y_{k, i} = 1$ if the $i$th speaker is dominant at the $k$th TF point, and $y_{k, i} = 0$ otherwise. Based on these definitions, we can obtain the estimated Wiener mask $\hat{m}_{f, n}^{(i)}$ for each $i$th speaker from input mixture $X$ through end-to-end training.

By using the same data sets as in Section \ref{sec:experiment}, we compared the speech separation performance of the proposed X-DC and the conventional methods including DANet \footnote{We implemented DANet by referring to the source code provided by the authors of \cite{Chen2017}: \url{https://github.com/naplab/DANet}.}. Aside from the hyperparameter settings for DANet, we used the same experimental settings as in Section \ref{sec:experiment}. The detailed numerical settings for DANet are as follows: 
\begin{itemize}
\item For any number of speakers in the training data set, we set the number of hidden cells in each BLSTM layers at $300$, the number of the BLSTM layers at $3$, the learning rate at $10^{-5}$, and the number of epochs for training at $T = 800$.
\item When there are two speakers in the training data set, we set the dimension of the embedding space at $D = 70$. Under this setting, the number of learnable parameters in DANet is $10,747,760$.
\item When there are four speakers in the training data set, we set $D = 70$. Under this setting, the number of learnable parameters in DANet is $10,747,760$.
\item When there are eight speakers in the training data set, we set $D = 20$. Under this setting, the number of learnable parameters in DANet is $6,901,360$.
\end{itemize}
The above hyperparameter settings of $D$ and $T$ were chosen by the hold-out validation with $66$ validation data sets.

\begin{figure*}[p]
\centering
\includegraphics[width=\hsize]{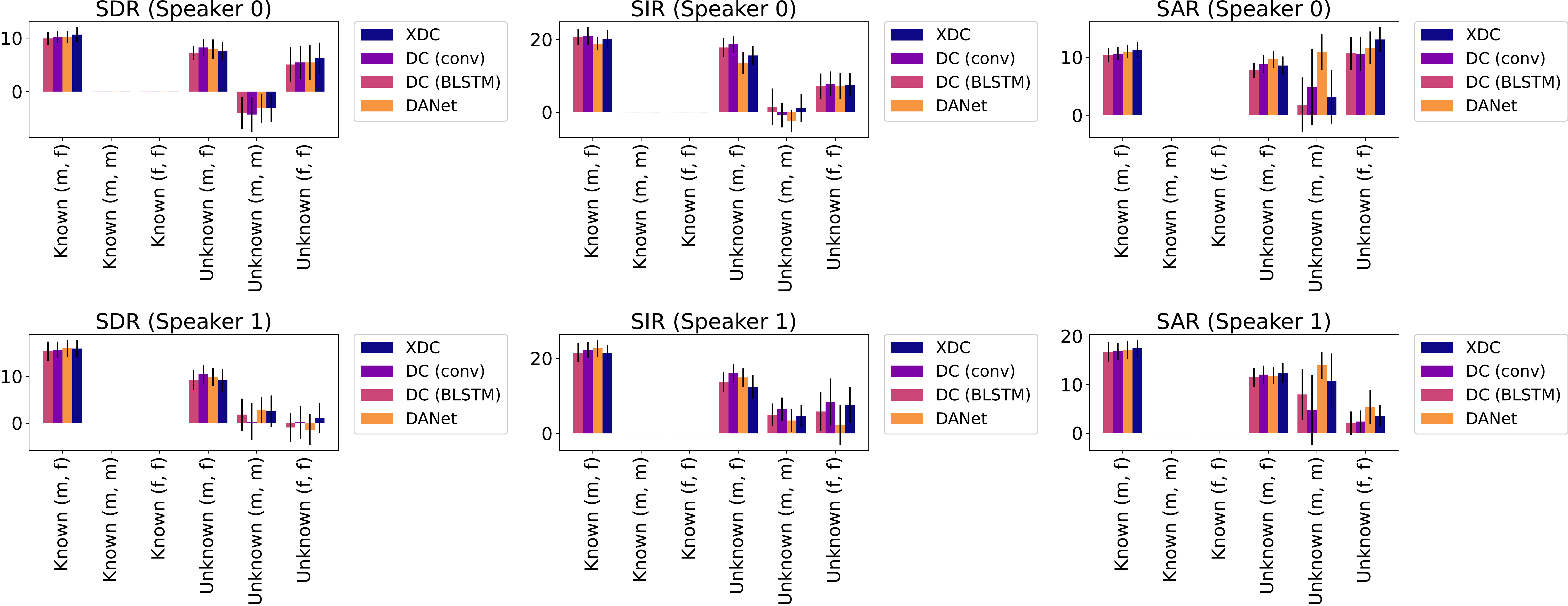}\vspace{-2mm}
\caption{Comparison of speech separation performance between the proposed X-DC, conv-DC, BLSTM-DC, and DANet (\textbf{$2$ speakers} in the training data set).}\vspace{5mm}
\label{fig:SDR_danet_2}
\includegraphics[width=\hsize]{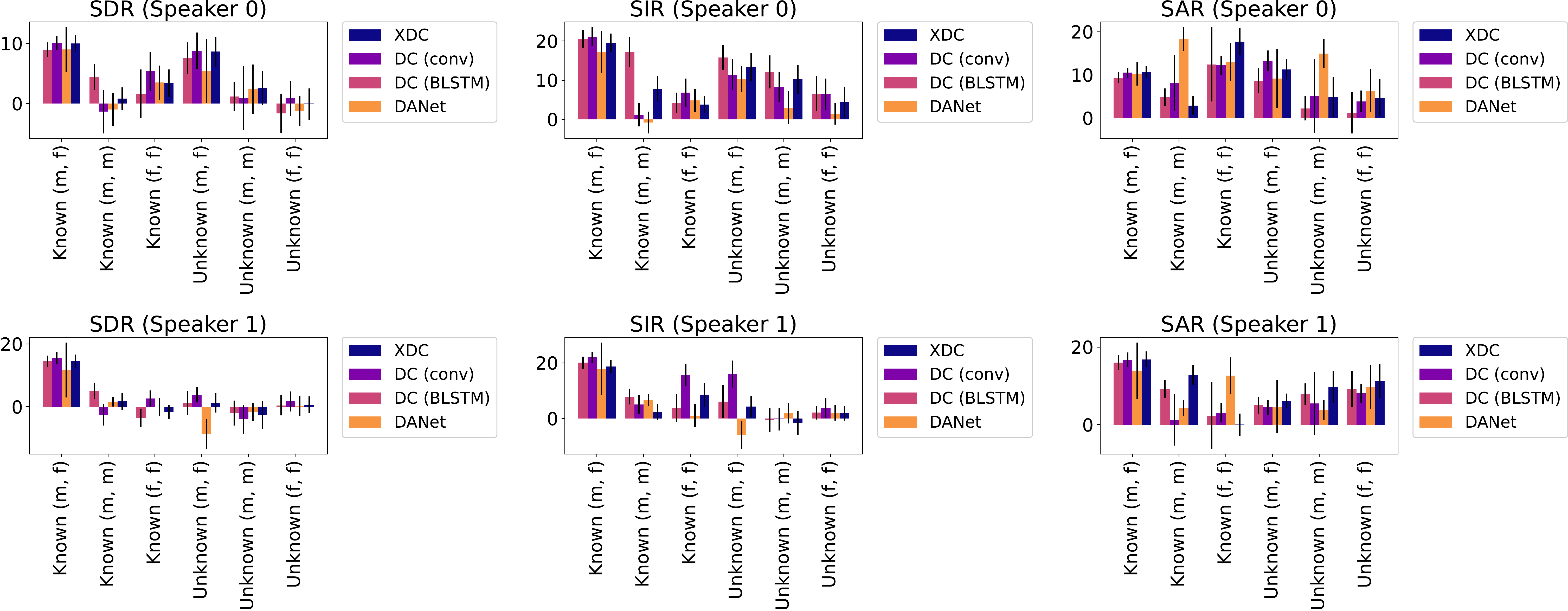}\vspace{-2mm}
\caption{Comparison of speech separation performance between the proposed X-DC, conv-DC, BLSTM-DC, and DANet (\textbf{$4$ speakers} in the training data set).}\vspace{5mm}
\label{fig:SDR_danet_4}
\includegraphics[width=\hsize]{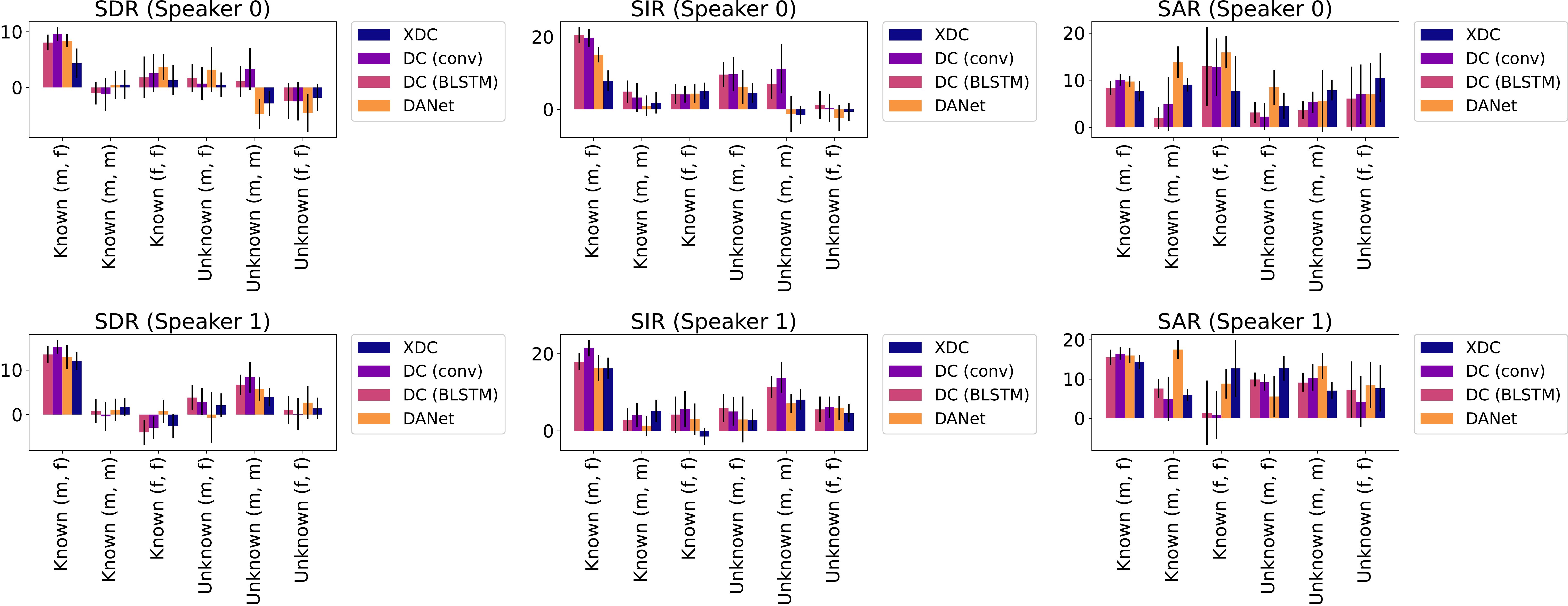}\vspace{-2mm}
\caption{Comparison of speech separation performance between the proposed X-DC, conv-DC, BLSTM-DC, and DANet (\textbf{$8$ speakers} in the training data set).}
\label{fig:SDR_danet_8}
\end{figure*}

Figures \ref{fig:SDR_danet_2}, \ref{fig:SDR_danet_4}, and \ref{fig:SDR_danet_8}, show the speech separation performance of the proposed and conventional models including DANet, where the training data set contains $2$, $4$, and $8$ different speakers, respectively. From these results, we see that DANet achieved as high speech separation performance as the other DC methods in most settings.

%---

\end{appendices}

\clearpage
\bibliographystyle{abbrv}
\bibliography{paper}

\begin{thebibliography}{10}

\bibitem{Bach2006}
F.~R. Bach and M.~I. Jordan.
\newblock Learning spectral clustering, with application to speech separation.
\newblock {\em Journal of Machine Learning Research}, 7:1963--2001, 2006.

\bibitem{Berry2007}
M.~W. Berry, M.~Browne, A.~N. Langville, V.~P. Pauca, and R.~J. Plemmons.
\newblock Algorithms and applications for approximate nonnegative matrix
  factorization.
\newblock {\em Computational Statistics \& Data Analysis}, 52(1):155--173,
  2007.

\bibitem{Chen2017}
Z.~Chen, Y.~Luo, and N.~Mesgarani.
\newblock Deep attractor network for single-microphone speaker separation.
\newblock In {\em 2017 IEEE International Conference on Acoustics, Speech and
  Signal Processing}, pages 246--250, 2017.

\bibitem{Cichocki2006}
A.~Cichocki, R.~Zdunek, and S.~Amari.
\newblock New algorithms for non-negative matrix factorization in applications
  to blind source separation.
\newblock In {\em 2006 IEEE International Conference on Acoustics, Speech and
  Signal Processing}, 2006.

\bibitem{Duan2008}
Z.~Duan, Y.~Zhang, C.~Zhang, and Z.~Shi.
\newblock Unsupervised single-channel music source separation by average
  harmonic structure modeling.
\newblock {\em IEEE Transactions on Audio, Speech, and Language Processing},
  16(4):766--778, 2008.

\bibitem{Fevotte2009}
C.~F\'{e}votte, N.~Bertin, and J.-L. Durrieu.
\newblock Nonnegative matrix factorization with the {I}takura-{S}aito
  divergence: With application to music analysis.
\newblock {\em Neural Computation}, 21(3):793--830, 2009.

\bibitem{Han2011}
K.~Han and D.~Wang.
\newblock An {SVM} based classification approach to speech separation.
\newblock In {\em 2011 IEEE International Conference on Acoustics, Speech and
  Signal Processing}, pages 4632--4635, 2011.

\bibitem{Hershey2016}
J.~R. Hershey, Z.~Chen, J.~{Le Roux}, and S.~Watanabe.
\newblock Deep clustering: Discriminative embeddings for segmentation and
  separation.
\newblock In {\em 2016 IEEE International Conference on Acoustics, Speech and
  Signal Processing}, pages 31--35, 2016.

\bibitem{Hershey2014}
J.~R. Hershey, J.~{Le Roux}, and F.~Weninger.
\newblock Deep unfolding: Model-based inspiration of novel deep architectures.
\newblock arXiv:1409.2574, 2014.

\bibitem{Kameoka2009}
H.~Kameoka, T.~Nakatani, and T.~Yoshioka.
\newblock Robust speech dereverberation based on non-negativity and sparse
  nature of speech spectrograms.
\newblock In {\em 2009 IEEE International Conference on Acoustics, Speech and
  Signal Processing}, pages 45--48, 2009.

\bibitem{Kingma2015}
D.~P. Kingma and J.~Ba.
\newblock {A}dam: A method for stochastic optimization.
\newblock In {\em International Conference on Learning Representations}, 2015.

\bibitem{arctic2004}
J.~Kominek and A.~W. Black.
\newblock The {CMU} {A}rctic speech databases.
\newblock In {\em 5th ISCA Speech Synthesis Workshop}, pages 223--224, 2004.

\bibitem{Roux2015}
J.~{Le Roux}, J.~R. Hershey, and F.~Weninger.
\newblock Deep {NMF} for speech separation.
\newblock In {\em 2015 IEEE International Conference on Acoustics, Speech and
  Signal Processing}, pages 66--70, 2015.

\bibitem{Lee2001}
D.~Lee and H.~S. Seung.
\newblock Algorithms for non-negative matrix factorization.
\newblock In {\em Advances in Neural Information Processing Systems}, pages
  556--562, 2001.

\bibitem{Lee1999}
D.~D. Lee and H.~S. Seung.
\newblock Learning the parts of objects by non-negative matrix factorization.
\newblock {\em Nature}, 401:788--791, 1999.

\bibitem{Li2018}
L.~Li and H.~Kameoka.
\newblock Deep clustering with gated convolutional networks.
\newblock In {\em 2018 IEEE International Conference on Acoustics, Speech and
  Signal Processing}, pages 16--20, 2018.

\bibitem{Luo2017}
Y.~Luo, Z.~Chen, J.~R. Hershey, J.~{Le Roux}, and N.~Mesgarani.
\newblock Deep clustering and conventional networks for music separation:
  Stronger together.
\newblock In {\em 2017 IEEE International Conference on Acoustics, Speech and
  Signal Processing}, pages 61--65, 2017.

\bibitem{Paatero1997}
P.~Paatero.
\newblock Least squares formulation of robust non-negative factor analysis.
\newblock {\em Chemometrics and Intelligent Laboratory Systems}, 37(1):23--35,
  1997.

\bibitem{Rabiee2012}
A.~Rabiee, S.~Setayeshi, and S.-Y. Lee.
\newblock A harmonic-based biologically inspired approach to monaural speech
  separation.
\newblock {\em IEEE Signal Processing Letters}, 19(9):559--562, 2012.

\bibitem{Schmidt2006nmfd}
M.~N. Schmidt and M.~M{\o}rup.
\newblock Nonnegative matrix factor 2-{D} deconvolution for blind single
  channel source separation.
\newblock In {\em Independent Component Analysis and Blind Signal Separation},
  pages 700--707, 2006.

\bibitem{Schmidt2006nmf}
M.~N. Schmidt and R.~K. Olsson.
\newblock Single-channel speech separation using sparse non-negative matrix
  factorization.
\newblock In {\em Ninth International Conference on Spoken Language
  Processing}, pages 2614--2617, 2006.

\bibitem{Smaragdis2004nmf}
P.~Smaragdis.
\newblock Discovering auditory objects through non-negativity constraints.
\newblock In {\em Statistical and Perceptual Audio Processing}, 2004.

\bibitem{Smaragdis2004nmfd}
P.~Smaragdis.
\newblock Non-negative matrix factor deconvolution; extraction of multiple
  sound sources from monophonic inputs.
\newblock In {\em Independent Component Analysis and Blind Signal Separation},
  pages 494--499, 2004.

\bibitem{Smaragdis2007}
P.~Smaragdis.
\newblock Convolutive speech bases and their application to supervised speech
  separation.
\newblock {\em IEEE Transactions on Audio, Speech, and Language Processing},
  15(1):1--12, 2007.

\bibitem{Vincent2006}
E.~Vincent, R.~Gribonval, and C.~F\'{e}votte.
\newblock Performance measurement in blind audio source separation.
\newblock {\em IEEE Transactions on Audio, Speech, and Language Processing},
  14(4):1462--1469, 2006.

\bibitem{Virtanen2007}
T.~Virtanen.
\newblock Monaural sound source separation by nonnegative matrix factorization
  with temporal continuity and sparseness criteria.
\newblock {\em IEEE Transactions on Audio, Speech, and Language Processing},
  15(3):1066--1074, 2007.

\bibitem{Wang2018b}
Z.-Q. Wang, J.~{Le Roux}, and J.~R. Hershey.
\newblock Multi-channel deep clustering: Discriminative spectral and spatial
  embeddings for speaker-independent speech separation.
\newblock In {\em 2018 IEEE International Conference on Acoustics, Speech and
  Signal Processing}, pages 1--5, 2018.

\bibitem{Wang2018a}
Z.-Q. Wang, J.~{Le Roux}, D.~Wang, and J.~R. Hershey.
\newblock End-to-end speech separation with unfolded iterative phase
  reconstruction.
\newblock In {\em Proceeding of Interspeech 2018}, pages 2708--2712, 2018.

\bibitem{Wisdom2017}
S.~Wisdom, T.~Powers, J.~Pitton, and L.~Atlas.
\newblock Deep recurrent {NMF} for speech separation by unfolding iterative
  thresholding.
\newblock In {\em 2017 IEEE Workshop on Applications of Signal Processing to
  Audio and Acoustics}, pages 254--258, 2017.

\bibitem{Xu2014}
Y.~Xu, J.~Du, L.-R. Dai, and C.-H. Lee.
\newblock An experimental study on speech enhancement based on deep neural
  networks.
\newblock {\em IEEE Signal Processing Letters}, 21(1):65--68, 2014.

\bibitem{Yu2017}
D.~Yu, M.~K. k, Z.-H. Tan, and J.~Jensen.
\newblock Permutation invariant training of deep models for speaker-independent
  multi-talker speech separation.
\newblock In {\em 2017 IEEE International Conference on Acoustics, Speech and
  Signal Processing}, pages 241--245, 2017.

\bibitem{arctic2018}
G.~Zhao, S.~Sonsaat, A.~Silpachai, I.~Lucic, E.~Chukharev-Hudilainen, J.~Levis,
  and R.~Gutierrez-Osuna.
\newblock {L2}-{ARCTIC}: A non-native {E}nglish speech corpus.
\newblock In {\em Proceeding of Interspeech 2018}, pages 2783--2787, 2018.

\end{thebibliography}

\end{document}